\documentclass[notitlepage,11pt,amsmath,preprintnumbers,superscriptaddress,nofootinbib,aps] 
{revtex4-1} 

\usepackage{amssymb,url,bm}
\usepackage{amsfonts,multirow}
\usepackage{pifont}
\usepackage{amsmath}
\usepackage{graphicx}
\usepackage{hyperref}
\usepackage{color}
\usepackage{xcolor,colortbl}
\usepackage{array,mathtools,amssymb,booktabs}
\usepackage{setspace}

\renewcommand{\thesection}{{\bf \Roman{section}}}

\long\def\del #1 \enddel { }

\newcommand{\tab}[1]{Tab.~\ref{#1}}

\newcommand{\sct}[1]{Sect.\,\ref{#1}}
\newcommand{\scts}[1]{Sects.\,\ref{#1}}

\definecolor{Gray}{gray}{0.85}
\definecolor{LightGray}{gray}{0.93}
\definecolor{LightGreen}{rgb}{0.88, 1, 0.88}
\definecolor{LightCyan}{rgb}{0.88,1,1}
\definecolor{LightRed}{rgb}{1, 0.85, 0.85}
\definecolor{LightYellow}{rgb}{1, 1, 0.85}
\definecolor{Yellow}{rgb}{1,1,0.05}
\definecolor{LightBlue}{rgb}{0.87, 0.94, 1}
\definecolor{white}{gray}{1}

\newcounter{notecount}

\newcolumntype{G}{>{\columncolor{LightGray}}c}
\newcolumntype{C}{>{$}c<{$}}
\newcolumntype{?}{!{\vrule width 1pt}}
\newcolumntype{`}{!{\vrule width 2pt}}

\AtBeginDocument{
\heavyrulewidth=.16em
\lightrulewidth=.1em
\cmidrulewidth=.03em
\belowrulesep=.4ex
\belowbottomsep=0pt
\aboverulesep=.4ex
\abovetopsep=0pt
\cmidrulesep=\doublerulesep
\cmidrulekern=.5em
\defaultaddspace=.5em
}

\def\beq{\begin{equation}}
\def\eeq{\end{equation}}

\def\bea{\arraycolsep .1em \begin{eqnarray}}
\def\eea{\end{eqnarray}}
\def\Tr{{\rm Tr}}

\def\eq#1{(\ref{#1})}

\def\s0#1#2{\mbox{\small{$ \frac{#1}{#2} $}}}
\def\0#1#2{\frac{#1}{#2}}

\def\grgl{\:\hbox to -0.2pt{\lower2.5pt\hbox{$\sim$}\hss}{\raise3pt\hbox{$>$}}\:}
\def\klgl{\:\hbox to -0.2pt{\lower2.5pt\hbox{$\sim$}\hss}{\raise3pt\hbox{$<$}}\:}

\makeatletter
    \def\CT@@do@color{%
      \global\let\CT@do@color\relax
            \@tempdima\wd\z@
            \advance\@tempdima\@tempdimb
            \advance\@tempdima\@tempdimc
    \advance\@tempdimb\tabcolsep
    \advance\@tempdimc\tabcolsep
    \advance\@tempdima2\tabcolsep
            \kern-\@tempdimb
            \leaders\vrule
                    \hskip\@tempdima\@plus  1fill
            \kern-\@tempdimc
            \hskip-\wd\z@ \@plus -1fill }

\begin{document}
\title{All Heat Kernel Coefficients on the Sphere}
\title{Heat Kernels Coefficients for Quantum Gravity  on the Sphere}
\title{Heat kernel coefficients on the sphere in any dimension}
\author{Yannick Kluth}
\author{Daniel F. Litim}
\affiliation{Department of Physics and Astronomy
University of Sussex, Brighton, BN1 9QH, U.K.}
\begin{abstract}
 We derive all heat kernel coefficients for Laplacians acting on scalars, vectors, and tensors 
 on fully symmetric spaces, in any dimension. Final expressions are  easy to evaluate and implement, and  confirmed independently using spectral sums and the Euler-Maclaurin formula. We also obtain the Green's function for Laplacians acting on transverse traceless tensors in any dimension, and new integral representations for heat kernels using known eigenvalue spectra of Laplacians.  Applications to quantum gravity and the functional renormalisation group, and other, are indicated.
\end{abstract}

\maketitle

\begin{spacing}{1}
\tableofcontents
\end{spacing}

\section{\bf Introduction}

Heat kernel techniques are well-established tools in both theoretical physics and mathematics
\cite{Fock:1937dy,Schwinger:1951nm,DeWitt:1967yk,DeWitt:1967ub,DeWitt:1967uc}. They  aim at the computation of traces of differential operators, and have  a wide range of applications  covering  
  fluctuations of quantum fields on curved space-times, ultraviolet divergences and effective actions, spectral functions, quantum anomalies, the Casimir effect, quantum gravity, and more
  \cite{Vassilevich:2003xt,Bordag:2001qi,Barvinsky:1985an,Avramidi:2000bm,Birrell:1982ix,Fradkin:1983mq,Camporesi:1990wm}.
The central  idea is to express a certain Green's function as an integral over the so called proper time parameter which satisfies the heat equation. The integrand of this method, the heat kernel, is then a functional of the background metric. While for some special backgrounds it can be calculated exactly, it is not possible to solve it for general manifolds.  Approximation schemes have been introduced including the seminal Schwinger-DeWitt expansion giving rise to the heat kernel coefficients. This  asymptotic expansion at early  proper time works well for small space-time separations, which makes it a convenient tool to study short distance  divergences in quantum field theory. On general manifolds, the first six heat kernel coefficients  are known  \cite{DeWitt:1965jb,Gilkey:1975iq,Christensen:1976vb,Christensen:1978yd,Amsterdamski:1989bt,Avramidi:1989ik,Avramidi:1990je,Avramidi:1990ug,vandeVen:1997pf}. Calculations of  heat kernel coefficients are greatly simplified on specific  manifolds such as maximally symmetric backgrounds where they can be obtained from  Green's function   
\cite{Avramidi:2000bm}.

Recent applications of heat kernels and spectral sums on dS or AdS  spaces cover a wide range of activities such as   tests of the AdS/CFT correspondence for conformal higher spin models  \cite{Giombi:2013fka,Giombi:2014yra}, derivations of effective actions in supergravity on curved backgrounds \cite{David:2009xg}, and studies of trace anomalies from quantum mechanical path integrals \cite{Bastianelli:2017wsy}, and more.  In the context of quantum gravity,  heat kernels  appear prominently in renormalisation group studies of Hořava gravity \cite{Barvinsky:2017mal}, and in tests of the asymptotic safety conjecture  \cite{Weinberg:1980gg,Litim:2008tt},
 In the latter, specifically, heat kernels on maximally symmetric or Einstein spaces are  used to find beta functions  for  gravitational couplings  \cite{Reuter:1996cp,Souma:1999at,Lauscher:2002sq,Litim:2003vp,Fischer:2006fz,Machado:2007ea,Codello:2008vh,Litim:2008tt, Benedetti:2009rx,Niedermaier:2009zz,Benedetti:2012dx,Christiansen:2012rx,Falls:2013bv,Falls:2014tra,Falls:2017lst,Falls:2018ylp}. Intriguingly, results for fixed points and  scaling exponents show  that  the canonical mass dimension of  couplings  remains a good ordering principle, with asymptotically safe quantum gravity becoming ``as Gaussian as it gets''  \cite{Falls:2013bv,Falls:2014tra,Falls:2017lst,Falls:2018ylp}. Further technical choices \cite{Litim:2001up,Litim:2003vp} are commonly adopted to reduce the required  heat kernel coefficients on spheres to a finite set \cite{Codello:2008vh}. In general, however, the flow of  couplings with increasing canonical mass dimension necessitates  the knowledge of an increasing number of heat kernel coefficients, 
 many more than presently available.

In this paper, we fill this gap in the literature and compute all heat kernel coefficients for scalars, transverse vectors, and transverse traceless symmetric tensors on  fully symmetric backgrounds with Euclidean signature and positive curvature. 
Our primary input are the known Green's functions for the Laplacian 
acting on  scalars
 \cite{Dowker:1975tf} 
and  transverse vectors  \cite{Tsamis:2006gj,Narain:2014oja}.  
In addition, we derive the Green's function of the Laplacian acting on transverse traceless tensors
to find closed expressions for all heat kernel coefficients on spheres, in any dimension. 
Final expressions are easy to evaluate and implement, and  confirmed  using spectral sums. We also find new  integral representations for heat kernels  using known eigenvalue spectra of Laplacians and the Euler-Maclaurin formula. Besides their general interest, our findings enable  new  tests of the asymptotic safety conjecture  without resorting to flat backgrounds or spectral sums and  approximations thereof.

The remainder of this paper is organised as follows. In Sect.~\ref{preliminaries} we recall 
the definition of heat kernel coefficients from asymptotic expansions (\sct{sec:intro}) and their usage to calculate functional traces such as in Wilson's renormalisation group  (\sct{sec:renormgroup}). 
Sect.~\ref{sec:results}  contains the main derivation of  heat kernel coefficients.
After an outline of the  method (\sct{sec:greentechnique}) 
we compute the heat kernels for scalars (\scts{sec:resscalar}), transverse vectors (\ref{sec:resvector}), transverse traceless tensors (\ref{sec:restensor}), and the first coefficients of the asymptotic heat kernel expansion for unconstrained fields  (\sct{sec:resunconstrained}). In \sct{sec:resspectralsum}, we derive heat kernels using spectral sums and the Euler-Maclaurin formula. After an outline of our methodology (\sct{SpectralSum}), we compute  heat kernels  for selected integer dimensions and confirm our previous findings (\sct{SpectralSumResults}). 
We also find new spectral integral representations for heat kernel coefficients (\sct{SpectralIntegrals}) including in general dimensions (\sct{sct:anacon}). 
Three appendices additionally provide  expansion coefficients (App~\ref{sec:vecstrucfunc}), heat kernel coefficients  in even dimensions (App.~\ref{sec:heatkerneleven}), and a derivation of heat kernels from spectral integrals (App.~\ref{app:scalarspec}).
In \sct{sec:Discussion} we close with a  discussion of results and   future applications. 

\section{\bf Preliminaries}\label{preliminaries}
In this section, we recall basic definitions for  heat kernel coefficients and their usage in the context of Wilson's renormalisation group. Ultimately, we aim at finding the
heat kernel expansion for different fields on the sphere using their corresponding Green's functions. Thus, from now on 
we focus on  the specific case of a fully symmetric background manifold, even though some of the considerations are more general. 

\subsection{Heat Kernel Coefficients}\label{sec:intro}
The heat kernel $U_E (t, x, y)$ is defined as the solution of the heat equation
\begin{equation}\label{heat}
	\frac{\partial U_E (t, x, y)}{\partial t} = ( \nabla^2 + E) U_E (t, x, y) \, ,
\end{equation}
subject to the initial condition
\begin{equation}\label{heat1}
	U_E (0, x, y) = \frac{\delta (x - y)}{\sqrt{g}} \, .
\end{equation}
Note that the kernel $U_E (t, x, y)$ may contain Lorentz indices if we consider vector or tensor degrees of freedom. Throughout this section these indices are suppressed. By definition, $U_E (t, x, y)$ has the  dimension of an inverse volume. Also, $t$ is the proper time parameter with mass dimension two, $-\nabla^2$ is the Laplacian, and $E$ an endomorphism. The formal solution of \eq{heat} is given by 
\beq
	U_E (t, x, y) = e^{t (\nabla^2 + E)} \, .
	\label{eqn:heatkernelexp}
\eeq
Due to the symmetries of the chosen background, $U_E (t, x, y)$ can only depend on the proper time $t$ and the distance between the points $x$ and $y$. Therefore, defining $\sigma$ to be half the square of the geodesic distance between $x$ and $y$, we may write
\beq
	U_E (t, x, y) = U_E (t, \sigma) \, .
\eeq
For early times we expand the heat kernel as an asymptotic series following the DeWitt ansatz
\begin{equation}
	U_E (t, \sigma) = \frac{\Delta^{1/2}}{(4 \pi t)^{d/2}} \exp \left\{ - \frac{\sigma}{2 t} \right\} \sum_{n = 0}^\infty \left[ \tilde{b}_{2n} (E, \sigma) t^{n} + \tilde{c}_{d+2n} (E, \sigma) t^{d/2+n} \right] \, ,
	\label{eqn:heatkernel}
\end{equation}
where $\Delta$ is the Van Fleck-Morette determinant
\beq
	\Delta = \frac{\det \left[ - \nabla_\mu^x \nabla_\nu^y \sigma (x, y) \right]}{g^{1/2} (x) g^{1/2} (y)} \, .
\eeq
By definition, the heat kernel coefficients $\tilde{b}_{m}$ and $\tilde{c}_{m}$ have canonical mass dimension ${m}$  in any dimension. Just as $U_E$, these coefficients may carry Lorentz indices which are suppressed here.

Note that the ansatz  \eq{eqn:heatkernel} seems slightly different from those used in the literature \cite{Rahmede:2008dwa,Percacci:2017fkn,Avramidi:2000bm,Falls:2017lst} through the appearance of the coefficients $\tilde{c}_{d+2n}$. These terms only arise for heat kernels of constrained fields, such as transverse vector and transverse traceless tensor fields, and are related to the exclusion of lowest modes \cite{Christensen:1979iy,Gibbons:1978ji,Gibbons:1978ac,Fradkin:1983mq}. 
Excluded modes always produce contributions of the form $\exp (\alpha R t)$ which invariably give rise to terms with positive integer powers of the proper-time parameter when expanded for small times. For even dimensions, the terms $\tilde{c}_{d+2n}$ could be combined with the $\tilde{b}_{d+2n}$ coefficients of the same mass dimension. In odd dimensions, however, all $\tilde{c}_{d+2n}$ coefficients have  mass dimensions different from all $\tilde{b}_{2n}$ coefficients, and cannot be combined into a single coefficient. Hence, for the sake of generality, and given their distinctly different origins, we keep these coefficients separate for now. The coefficients $\tilde{c}_{d+2n}$ vanish for unconstrained fields as we will show a posteriori (see \sct{sec:resunconstrained}).

Since we are ultimately interested in the trace of the heat kernel, we only need the coincidence limit of $U_E (t, \sigma)$. Using \eq{eqn:heatkernel}, we define the coincidence limit of the heat kernel coefficients, $\tilde{b}_{2n} (E)$ and $\tilde{c}_{d + 2n} (E)$, for given endomorphism $E$ as
\begin{equation}
	\tilde{b}_{2n} (E) = \tilde{b}_{2n} (E, 0) \, , \qquad \tilde{c}_{d + 2n} (E) = \tilde{c}_{d+2n} (E, 0) \, .
\end{equation}
Then, the trace of the heat kernel is given by
\begin{equation}
	\text{Tr}_s U_E (t, \sigma) = \frac{1}{(4 \pi t)^{d/2}} \sum_{n = 0}^\infty \left[ \Tr_s \, [ \tilde{b}_{2n} (E) ] \, t^{n} + \Tr_s \, [ \tilde{c}_{d + 2n} (E) ] \, t^{d/2 + n} \right] \, ,
\end{equation}
where the trace acts on the coordinate dependence as well as any Lorentz indices carried by $\tilde{b}_{2n}$ and $\tilde{c}_{d + 2n}$. Further, the index $s$ denotes the spin of the field w.r.t. which the trace is acting on. Since the heat kernel coefficients are coordinate independent on a fully symmetric background, we may define
\beq\label{translate}
	b_n^{(s)} (E) = \frac{1}{\text{Vol}} \Tr_s [ \tilde{b}_n (E) ] \, , \qquad c_n^{(s)} (E) = \frac{1}{\text{Vol}} \Tr_s [ \tilde{c}_n (E) ] \, ,
\eeq
in which the volume of the $d$-dimensional sphere is
\beq\label{vol}
	\text{Vol} = \frac{2 \pi^{(d + 1)/2}}{\Gamma \left( \frac{d + 1}{2} \right)} \left( \frac{d (d - 1)}{R} \right)^{d/2} \, 
\eeq
and $R$ denotes the Ricci scalar curvature.
This allows us to write
\beq\label{TrsU}
	\text{Tr}_s U_E (t, \sigma) =  \frac{\text{Vol}}{(4 \pi t)^{d/2}} \sum_{n = 0}^\infty \left[ b_{2n}^{(s)} (E) t^{n} + c_{d + 2n}^{(s)} (E) t^{d/2 + n} \right] \,.
\eeq
Finally, we notice that the heat kernel for a given endomorphism $\overline{E}$ is related to that for any other endomorphism $E$ by
\begin{equation}
	U_E (t, \sigma) = e^{t (E - \overline{E})} U_{\overline{E}} (t, \sigma) \,,
\end{equation}
assuming that the endomorphism commutes with the covariant derivative on the sphere. 
This relation implies that the corresponding heat kernel coefficients are related by
\begin{equation}
	\begin{split}
		b_{2n}^{(s)} (E) =& \sum_{k = 0}^n \frac{\left( E - \overline{E} \right)^k}{k!} b_{2 (n - k)}^{(s)} (\overline{E}) \, , \\
		c_{d + 2n}^{(s)} (E) =& \sum_{k = 0}^n \frac{\left( E - \overline{E} \right)^k}{k!} c_{d + 2 (n - k)}^{(s)} (\overline{E}) \, ,
	\end{split}
	\label{eqn:heatkgeneralendo}
\end{equation}
which serves as a definition of heat kernel coefficients for arbitrary endomorphisms.

\subsection{Renormalisation Group}\label{sec:renormgroup}
An important area for the application of heat kernels is Wilson's  (functional) renormalisation group. The technique amounts to the introduction of an infrared momentum cutoff $k$ into the path integral definition of quantum or statistical field theory, which induces a  scale-dependence $k\partial_k$  in the form of  an exact functional flow for  the effective  action $\Gamma_k$ (see \cite{Litim:1998nf,Berges:2000ew,Litim:2011cp} for  reviews),
\begin{equation}
	k\partial_k \Gamma_k = \frac{1}{2} \text{Tr} \left\{ \Big( k\partial_k \mathcal{R}_k \Big) \left( \Gamma_k^{(2)} + \mathcal{R}_k \right)^{-1} \right\} \,.
	\label{eqn:wetterich}
\end{equation}
Here, $\Gamma_k^{(2)}$ denotes the second variation of $\Gamma_k$. The function $\mathcal{R}_k$ denotes the Wilsonian IR regulator, chosen such that  $\Gamma_k$ interpolates between the microscopic theory ($1/k\to 0$) and the full quantum effective action $(k\to 0)$, see \cite{Litim:2001up}.  At weak coupling, iterative solutions generate  perturbation theory to all  loop orders  \cite{Litim:2001ky,Litim:2002xm}.  At strong coupling,  non-perturbative  approximations such as the derivative expansion, vertex expansions, or mixtures thereof are available  \cite{Litim:1998nf,Pawlowski:2005xe,Reuter:2012id}. The stability and convergence of approximations can be controlled as well  \cite{Litim:2001fd,Litim:2010tt,Balog:2019rrg}.

Our main point here relates to the operator trace in \eq{eqn:wetterich}, which for many applications can be  evaluated on flat Euclidean backgrounds. For quantum field theories on  curved backgrounds, or for studies of fully-fledged quantum gravity, it is often convenient  to evaluate  the operator trace  on suitably chosen  non-flat backgrounds 
\cite{Floreanini:1995aj,Reuter:1996cp,Dou:1997fg,Lauscher:2002sq}
also using the background field method \cite{Freire:2000bq} and optimised cutoffs \cite{Litim:2000ci,Litim:2001up,Litim:2003vp,Litim:2006ag}. In quantum gravity, this has enabled advanced tests of the asymptotic safety conjecture on spheres \cite{Litim:2003vp,Fischer:2006fz,Codello:2008vh,Falls:2013bv,Falls:2014tra,Falls:2016wsa,Falls:2018ylp}.
Further applications of heat kernels include  flow equations on Einstein  \cite{Benedetti:2009gn} or hyperbolic spaces \cite{Falls:2016msz},   critical fields on curved  backgrounds \cite{Benedetti:2014gja}, low energy effective actions  \cite{Codello:2015oqa}, and proper-time flows \cite{Bonanno:2004sy}.

To see how heat kernel coefficients enter in this methodology we note that typical contributions on the right-hand side of \eq{eqn:wetterich} are given by traces of operators in the form $( k\partial_k \mathcal{R}_k ) ( \Gamma_k^{(2)} + \mathcal{R}_k )^{-1}$. After inserting the second variation matrix and choosing a regulator the integrand  can  be represented as a (matrix-valued) function $W (-\nabla^2)$ of the Laplacian. Using the Laplace anti-transformation
\beq\nonumber
	W (z) = \int_0^\infty \text{d} t \, \widetilde{W} (t) e^{-t z} \, ,
\eeq
we may then express the desired trace as 
\beq\label{trace}
	\Tr \, W ( -\nabla^2 ) = \int_0^\infty \text{d} t \, \widetilde{W} (t) \,\Tr \, e^{t \nabla^2} \, .
\eeq
Crucially, the trace $\Tr \, e^{t \nabla^2}$ appearing on the right-hand side is the trace of the heat kernel with vanishing endomorphism  \eq{eqn:heatkernelexp}. Thus, the early time expansion  \eq{eqn:heatkernel} allows us to evaluate the trace  \eq{trace} in terms of the heat kernel coefficients $b_{2n}$ and $c_{d + 2n}$ via
\beq\label{TrW}
	\Tr \, W ( -\nabla^2 ) = \frac{\text{Vol}}{(4 \pi)^{d/2}} \sum_{n = 0}^\infty \left[ b_{2n} (0) \int_0^\infty \text{d} t \, t^{n - d/2} \, \widetilde{W} (t) + c_{d + 2n} (0) \int_0^\infty \text{d} t \, t^{n} \, \widetilde{W} (t) \right] \, .
\eeq
We conclude that the heat kernel coefficients $b_{2n}(0)$ and $c_{d + 2n}(0)$ are key inputs for Wilsonian flows  on maximally symmetric backgrounds. The calculation of all  heat kernel coefficients $b_{2n}$ and $c_{d + 2n}$  $(n\ge 0)$ on spheres in arbitrary dimension is the topic of the following sections.

\section{\bf Heat Kernels from Green's Functions}\label{sec:results}
In this section, we  find Green's functions for scalars,  transverse vectors, and transverse traceless tensors and use these to extract heat kernel coefficients for Laplacians on spheres in any dimension. We also give results for heat kernels of unconstrained vectors and tensors.

\subsection{Green's Function Technique}\label{sec:greentechnique}
Our methodology largely follows  Avramidi \cite{Avramidi:2000bm} and starts by  noting that the heat kernel defined in \eq{eqn:heatkernel} can be connected to a Green's function using the Schwinger-DeWitt representation
\begin{equation}\label{Gsigma}
	G (\sigma) = \int_0^\infty \text{d} t \, e^{-t m^2} U_E (t, \sigma) \, .
\end{equation}
By definition, $G (\sigma)$ has canonical mass dimension $M^{d-2}$. Using \eq{heat} with \eq{heat1} it is straightforward to show that $G (\sigma)$ in \eq{Gsigma} is a Green's function for the differential operator $(- \nabla^2 + m^2 - E)$.
In the coincidence limit $\sigma = 0$, we can write
\begin{equation}\label{G0}
	\frac{\Tr_s G (\sigma)}{\text{Vol}} = \frac{1}{(4 \pi)^{d/2}} \int_0^\infty \text{d} t \, e^{-t m^2} \sum_{n = 0}^\infty \left[ b_{2n}^{(s)} (E) t^{-d/2 + n} + c_{d + 2n}^{(s)} (E) t^n \right] \, .
\end{equation}
Note that the trace on the left-hand side effectively only takes the trace of the tensor structure of $G (\sigma)$, the trace over the coordinate dependence drops out due to the volume factor in the denominator.  
Recalling the elementary definition of the $\Gamma$-function, $\Gamma(n+1)=\int_0^\infty {\rm d}s \, s^ne^{-s}$,
 and substituting $s=tm^2$,
the $t$-integration in  \eq{G0}  is performed term by term. Doing so, the Green's function 
takes the form of a large-$m$ expansion
\begin{equation}
	\frac{\Tr_s G (\sigma)}{\text{Vol}} = \frac{m^{d-2}}{(4 \pi)^{d/2}} \sum_{n = 0}^\infty \left[ \Gamma (n - \s0d2 + 1) \frac{b_{2n}^{(s)} (E) }{m^{2n}} +  \Gamma(n + 1) \frac{c_{d + 2n}^{(s)} (E)}{m^{d +2n}} \right] \,.
	\label{eqn:heatreadoff}
\end{equation}
Hence, by calculating the large-$m$ expansion of the Green's function at its coincidence limit, we may read off the heat kernel coefficients as the corresponding Taylor coefficients. This fact is exploited below to calculate all heat kernel coefficients for scalars, transverse vectors, and transverse traceless symmetric tensors.

\subsection{Scalars}\label{sec:resscalar}
We begin  with  the Green's function for scalar fields $G_Q (\sigma)$ to explain how the corresponding heat kernel coefficients are computed in practice. This follows closely the derivation given in \cite{Avramidi:2000bm}. The Green's function for scalar fields is the solution of the differential equation
\begin{equation}
	(- \nabla^2 + Q) G_Q (\sigma) = \frac{1}{\sqrt{g}} \delta (x - y) \, .
\end{equation}
In  \cite{Dowker:1975tf,Allen:1985wd,Miao:2011fc} it has been explained why solutions can be expressed in terms of a hypergeometric function,
\begin{equation}
	G_Q (\sigma) = \frac{\Gamma(a) \Gamma(b)}{\Gamma (d/2)} \frac{r^{2 - d}}{(4 \pi)^{d/2}} \, _2F_1 (a, b; c; z)
	\label{eqn:scalarGreen}
\eeq
with parameters  
\beq
	a = \frac{d - 1}{2} + \xi \, , \quad b = \frac{d - 1}{2} - \xi \, , \quad c = \frac{d}{2} \, , \quad \xi = \sqrt{\frac{(d - 1)^2}{4} - Q r^2} \, ,
	\label{eqn:scalarGreenABC}
\eeq
and
\beq
	z = \cos^2 \left( \sqrt{\frac{\sigma}{2 r^2}} \right) \, .
\end{equation}
In the latter, $r$ denotes the radius of the sphere which relates to the Ricci scalar curvature $R$ as
\begin{equation}
	\frac{R}{d (d - 1)} = r^{-2} \, .
	\label{r}
\end{equation}
To find the coincidence limit for \eq{eqn:scalarGreen}, we follow Avramidi \cite{Avramidi:2000bm} and exploit  a useful representation for the hypergeometric function  \cite{Gammacoincidence}
\begin{equation*}
	_2F_1 (a, b; c; 1) = \frac{\Gamma (c) \Gamma(c - a - b)}{ \Gamma(c - a) \Gamma (c - b)}
\end{equation*}
to arrive at
\begin{equation}\label{GQ}
	\begin{split}
		G_Q (0) 
		&= \frac{r^{2 - d}}{(4 \pi)^{d/2}} \Gamma \left( 1 - \frac{d}{2} \right) \frac{\Gamma \left( \frac{d - 1}{2} + \xi \right) \Gamma \left( \frac{d - 1}{2} - \xi \right)}{\Gamma \left( \frac{1}{2} + \xi \right) \Gamma \left( \frac{1}{2} - \xi \right)} \, .
	\end{split}
\end{equation}
The large mass expansion requires an expansion of the Gamma functions which can be done noting that \cite{Gammaexpand}
\begin{equation*}
	\ln \left[ \Gamma (\alpha + \xi) \right]= (\alpha + \xi - \012) \ln (\xi) - \xi + \frac{1}{2} \ln(2 \pi) + \sum_{n = 2}^\infty \frac{(-1)^n B_n (\alpha)}{n (n - 1) \xi^{n - 1}} \, ,
\end{equation*}
with $|\xi| \rightarrow \infty$, $|\text{ph} (\xi)| < \pi$, and $B_n (x)$ being the Bernoulli polynomials. Then, using $B_n (\s012) = 0$ for $n = $ odd, we find 
\begin{equation}
	\frac{\Gamma \left( \frac{d - 1}{2} + \xi \right) \Gamma \left( \frac{d - 1}{2} - \xi \right)}{\Gamma \left( \frac{1}{2} + \xi \right) \Gamma \left( \frac{1}{2} - \xi \right)} = \sum_{n = 0}^\infty \kappa_n (d) \left( - \xi^2 \right)^{d/2 - 1 - n}\,,
	\label{Gammasimplify}
\end{equation}
with the generating function for the coefficients $\kappa_n(d)$  given by
\begin{equation}\label{kappa}
	\exp{ \sum_{n = 1}^\infty \frac{(-1)^{n + 1}}{n (2n + 1)} B_{2n + 1} \left( \frac{d - 1}{2} \right) z^n } 
	=\sum_{n = 0}^\infty \kappa_n (d) z^n \,.
\end{equation}
Note that the expression \eq{Gammasimplify} is only valid for $\Im (\xi) \neq 0$. To proceed, we  split $Q$ into a mass part $m^2$ and an endomorphism part $E$ through $Q = m^2 - E$. The endomorphism part $E$ can be chosen such that $\xi$ becomes proportional to the mass $m$. Using \eq{eqn:scalarGreenABC} with \eq{r}, this requirement uniquely fixes the  endomorphism  $E=\overline{E}$ and $\xi$ to
\beq\label{EbarS}
\begin{array}{rcl}
		\overline{E} &=&\displaystyle \frac{1-d}{4d}R
		\,, \\[2ex]
		-\xi^2&=&\displaystyle d (d - 1) \frac{m^2}{R} \,,
	\end{array}
\end{equation}
and we get
\begin{equation}\label{FinalScalarExpansion}
	G_Q (0)  \Big|_{\overline{E}}
	= \frac{1}{(4 \pi)^{d/2}} \Gamma \left( 1 - \frac{d}{2} \right) \sum_{n = 0}^\infty \kappa_n (d)\,  r^{- 2 n} \, m^{d - 2 - 2 n} \, .
\end{equation}
For the  endomorphism \eq{EbarS} we can now read off the heat kernel coefficients $b_{2n}^{(0)}$ and $c_{d + 2n}^{(0)}$ by comparison with \eq{eqn:heatreadoff}. To distinguish them, we note that the coefficients $c_{d + 2n}^{(0)}$ enter with $d$-independent integer powers of $m$ in \eq{eqn:heatreadoff} while the coefficients $b_{2n}^{(0)}$ have $d$-dependent powers of $m$. Moreover, the coefficients $b_{2n}^{(0)}$ ($c_{d + 2n}^{(0)}$) are linear in  (independent of) the parameters $\kappa_i (d)$. This is because the $d$-dependent powers of $m$ originate from \eq{Gammasimplify} which is linear in the $\kappa_i (d)$. Exploiting this fact, we get from \eq{FinalScalarExpansion}
\begin{equation}\label{bcScalarBar}
	\begin{split}
		b_{2n} ^{(0)}\left( \overline{E} \right) &= \frac{\Gamma \left( 1 - \frac{d}{2} \right)}{\Gamma \left( 1 + n - \frac{d}{2} \right)} \left( \frac{R}{d (d - 1)} \right)^n \kappa_n (d) \, , \\
		c_{d + 2n} ^{(0)}\left( \overline{E}\right) &= 0 \, ,
	\end{split}
\end{equation}
with $\kappa_n(d)$ determined through \eq{kappa}. Moreover, with the help of  \eq{eqn:heatkgeneralendo}, we find the heat kernel coefficients for scalar fields and  arbitrary endomorphism $E$,
\begin{equation}
	\begin{split}
		b_{2n} ^{(0)}(E) =& \sum_{k = 0}^n \frac{\Gamma \left(1-\frac{d}{2}\right)}{k! \,\Gamma \left(1 + n - k - \frac{d}{2} \right)} \left(\frac{(d-1) R}{4 d}+E\right)^k \left(\frac{R}{(d-1) d}\right)^{n-k} \kappa _{n-k}(d) \, , \\
		c_{d + 2n} ^{(0)}(E) =& \, 0 \, .
	\end{split}
	\label{eqn:resscalar}
\end{equation}
Tab.~\ref{tab:scalarHK} summarises our results for the first few scalar heat kernel coefficients \eq{eqn:resscalar} for $E=0$ and a selection of integer dimensions. 

Finally, we note that the heat kernel expansion with  \eq{eqn:resscalar}  is asymptotic in even dimensions but has a finite radius of convergence  in odd  ones, owing to the propagator \eq{GQ} being non-finite only in even dimensions \cite{Avramidi:2000bm}. The result generalises to non-integer dimensions.

\aboverulesep = 0mm
\belowrulesep = 0mm
\begin{table}
	\addtolength{\tabcolsep}{12pt}
	\setlength{\extrarowheight}{8pt}
	\begin{tabular}{`c?c|c|c|c|c`}
	\toprule  
 \rowcolor{Yellow} & $\bm{d = 2}$ & $\bm{d = 3}$ & $\bm{d = 4}$ & $\bm{d = 5}$ & $\bm{d = 6}$  \\[1ex] \midrule
$\bm{b_{0}^{(0)}}$ & $1$ & $1$ & $1$ & $1$ & $1$ \\[1ex]
\rowcolor{LightGray} $\bm{b_{2}^{(0)}}$ & $\frac{1}{6}R$ & $\frac{1}{6}R$ & $\frac{1}{6}R$ & $\frac{1}{6}R$ & $\frac{1}{6}R$ \\[1ex]
$\bm{b_{4}^{(0)}}$ & $\frac{1}{60}R^2$ & $\frac{1}{72}R^2$ & $\frac{29}{2160}R^2$ & $\frac{1}{75}R^2$ & $\frac{1}{75}R^2$ \\[1ex]
\rowcolor{LightGray} $\bm{b_{6}^{(0)}}$ & $\frac{1}{630}R^3$ & $\frac{1}{1296}R^3$ & $\frac{37}{54432}R^3$ & $\frac{1}{1500}R^3$ & $\frac{1139}{1701000}R^3$ \\[1ex]
$\bm{b_{8}^{(0)}}$ & $\frac{1}{5040}R^4$ & $\frac{1}{31104}R^4$ & $\frac{149}{6531840}R^4$ & $\frac{1}{45000}R^4$ & $\frac{833}{36450000}R^4$ \\[1ex]
\rowcolor{LightGray} $\bm{b_{10}^{(0)}}$ & $\frac{1}{27720}R^5$ & $\frac{1}{933120}R^5$ & $\frac{179}{431101440}R^5$ & $\frac{1}{2250000}R^5$ & $\frac{137}{267300000}R^5$ \\[1ex]
 \bottomrule 
	\end{tabular}
	\caption{The scalar heat kernel coefficients for different integer dimensions and vanishing endomorphism.}
	\label{tab:scalarHK}
\end{table}

\subsection{Transverse Vectors}\label{sec:resvector}
\label{sec:vectors}
Next, we  determine the heat kernel coefficients for transverse vector fields. To that end, we  review the derivation of the transverse vector Green's function, closely following \cite{Tsamis:2006gj,Miao:2011fc}. We then exploit the result to derive all heat kernel coefficients for transverse vector fields on spheres. 

It is important to notice that the Green's function for transverse vector fields $G_{Q, \mu \nu'}^T (\sigma)$ fulfils a differential equation of the form
\begin{equation}
	(- \nabla^2 + Q) G_{Q, \mu \nu'}^T (\sigma) = \frac{g_{\mu \nu'}}{\sqrt{g}} \delta(x - y) + \text{longitudinal terms} \, .
	\label{eqn:vectordgl}
\end{equation}
The longitudinal terms ensure the transversality of the right-hand side. They can be derived by considering the full vector Green's function and splitting it into a transverse and a longitudinal part. As pointed out in \cite{Tsamis:2006gj}, neglecting these longitudinal terms can lead to inconsistent results for the Green's function.

To solve the differential equation for the transverse vector Green's function, we want to reduce it to a scalar function $S^T (\sigma)$, which is called the structure function. For this, the transverse vector projector
\begin{equation}
	\mathcal{P}_\mu^{\,\,\,\nu} = g_\mu^{\,\,\,\nu} \nabla^2 - \nabla^\nu \nabla_\mu \, ,
\end{equation}
is introduced. It fulfils the properties \cite{Tsamis:2006gj,Miao:2011fc}
\begin{equation}
	\nabla^\mu \mathcal{P}_\mu^{\,\,\,\nu} T_\nu = \mathcal{P}_\nu^{\,\,\,\mu} \left( \nabla_\mu S (\sigma) \right) = 0 \, , \quad [\nabla^2, \mathcal{P}_{\mu}^{\,\,\,\nu}] T_\nu = 0 \, , \quad \mathcal{P}_\mu^{\,\,\,\nu} \mathcal{P}_\nu^{\,\,\,\rho} T_\rho = \mathcal{P}_\mu^{\,\,\,\rho} \left( \nabla^2 - \frac{R}{d} \right) T_\rho \, ,
\end{equation}
for an arbitrary vector $T_\mu$. From this, it follow that acting with $\mathcal{P}^{\,\,\,\alpha}_\mu \mathcal{P}^{\,\,\,\beta'}_{\nu'}$ on any bi-tensor $T_{\alpha \beta'}$ gives rise to a bi-tensor which is transverse in both indices. Hence, it has the right properties to be the Green's function for transverse vector fields.
With this in mind, we make an ansatz for the transverse vector Green's function through
\begin{equation}
	G_{Q, \mu \nu'}^T (\sigma) = \mathcal{P}^{\,\,\,\alpha}_\mu \mathcal{P}^{\,\,\,\beta'}_{\nu'} \left( \mathcal{R}_{\alpha \beta'} S^T (\sigma) \right) \, .
\end{equation}
The bi-tensor $\mathcal{R}_{\alpha \beta'}$ is arbitrary and can be chosen to simplify the computations. One choice is given by \cite{Tsamis:2006gj,Miao:2011fc}
\begin{equation}
	\mathcal{R}_{\alpha \beta'} = - 2 r^2 \nabla_\alpha \nabla_{\beta'} \sin^2 \left( \frac{\sqrt{2 \sigma}}{2 r} \right) = g_{\alpha \beta'} + \frac{\sigma_\alpha \sigma_{\beta'}}{\sigma} \sin^2 \left( \frac{\sqrt{2 \sigma}}{2 r} \right)\, ,
\end{equation}
with $\sigma_\alpha = \nabla_\alpha \sigma$. The right-hand side of this equation can be derived using the relations in \eq{eqn:sigmaderivs} and \eq{eqn:ACfunct}. It fulfils the property
\begin{equation}
	\begin{gathered}
		\nabla^2 \left( \mathcal{R}_{\alpha \beta'} S (\sigma) \right) = \mathcal{R}_{\alpha \beta'} \left( \nabla^2 + \frac{R}{d (d - 1)} \right) S (\sigma) + \text{longitudinal} \, ,
	\end{gathered}
	\label{eqn:vectorprojectorprop}
\end{equation}
with $S(\sigma)$ being an arbitrary scalar. Since this bi-tensor is always contracted with transverse vector projectors, the longitudinal terms can be neglected.
Using the above structures and identities, and contracting  \eq{eqn:vectordgl} with  $\mathcal{P}^{\,\,\,\mu}_\rho \mathcal{P}^{\,\,\,\nu'}_{\sigma'}$ we  arrive at
\begin{equation*}
		\mathcal{P}^{\,\,\,\alpha}_\rho \mathcal{P}^{\,\,\,\beta'}_{\sigma'} \left[ \mathcal{R}_{\alpha \beta'} \left(- \nabla^2 + Q - \frac{R}{d(d - 1)} \right) \left( \nabla^2 + \frac{2 - d}{d (d - 1)} R \right)^2 S^T (\sigma) \right]
		=
		 \mathcal{P}^{\,\,\,\alpha}_\rho \mathcal{P}^{\,\,\,\beta'}_{\sigma'} \left[ \frac{\mathcal{R}_{\alpha \beta'}}{\sqrt{g}} \delta (x - y) \right] \, .
\end{equation*}
We observe that the structure function for the transverse vector Green's function obeys the differential equation
\begin{equation}
	\left(- \nabla^2 + Q - \frac{R}{d(d - 1)} \right) \left( \nabla^2 + \frac{2 - d}{d (d - 1)} R \right)^2 S^T (\sigma) = \frac{1}{\sqrt{g}} \delta (x - y) \,,
\end{equation}
which  can be solved with the help of the scalar Green's function  \cite{Tsamis:2006gj,Miao:2011fc}.  Its solution takes the  explicit form
\begin{equation}\label{eqn:vectorGreen}
		S^T (\sigma) = G_{\chi_1, \chi_2, \chi_2} (\sigma) \equiv - \frac{\partial}{\partial \chi_2} G_{\chi_1, \chi_2} (\sigma) \, ,
		\eeq
		where
		\beq\label{Gchi1chi2}
		G_{\chi_1, \chi_2} (\sigma) = \frac{G_{\chi_1} (\sigma) - G_{\chi_2} (\sigma) }{\chi_2 - \chi_1}\,,
		\eeq
		and $G_\chi (\sigma) $ given in \eq{eqn:scalarGreen}, and all of this evaluated at
		\beq
		\chi_1 = Q - \frac{R}{d (d - 1)} \,, \quad 
		\chi_2 = \frac{d-2}{d (d - 1)} R\,.
	\label{chi12}
\eeq
Using the result for the structure function, the transverse vector Green's function is now given by
\begin{equation}
	G_{Q, \mu \nu'}^T (\sigma) = \mathcal{P}_\mu^{\ \alpha} \mathcal{P}_{\nu'}^{\ \beta'} \left( \mathcal{R}_{\alpha \beta'} G_{\chi_1, \chi_2, \chi_2} (\sigma) \right) \, .
\end{equation}
To extract the heat kernel coefficients we need to contract this with $g^{\mu \nu'}$ and calculate the coincidence limit of the resulting expression. Covariant derivatives acting on $\sigma_\alpha$ can be computed using \cite{Narain:2014oja}
\begin{equation}
	\begin{split}
		\nabla_{\nu'} \sigma_\mu =& C(\sigma) \left[ g_{\mu \nu'} + \frac{1}{2 \sigma} \sigma_\mu \sigma_{\nu'} \right] + \frac{1}{2 \sigma} \sigma_\mu \sigma_{\nu'} \, , \\
		\nabla_{\nu} \sigma_\mu =& A(\sigma) \left[ g_{\mu \nu} - \frac{1}{2 \sigma} \sigma_\mu \sigma_{\nu} \right] + \frac{1}{2 \sigma} \sigma_\mu \sigma_{\nu} \, , \\
		\nabla_\mu g_{\alpha \beta'} =& - \frac{A (\sigma) + C(\sigma)}{2 \sigma} \left( g_{\mu \alpha} \sigma_{\beta'} + g_{\mu \beta'} \sigma_\alpha \right) \, ,
	\end{split}
	\label{eqn:sigmaderivs}
\end{equation}
with
\begin{equation}
	A (\sigma) = \sqrt{\frac{2 \sigma R}{d (d - 1)}} \cot \left( \sqrt{\frac{2 \sigma R}{d (d - 1)}} \right) \, , \qquad C (\sigma) = - \sqrt{\frac{2 \sigma R}{d (d - 1)}} \csc \left( \sqrt{\frac{2 \sigma R}{d (d - 1)}} \right) \, .
	\label{eqn:ACfunct}
\end{equation}
Acting with the transverse vector projectors can be simplified by performing a series expansion of $\mathcal{R}_{\alpha \beta'} G_{\chi_1 \chi_2} (\sigma)$ in the coincidence limit. Observing that $\sigma_\mu \sigma^\mu = 2 \sigma$ and that $\sigma_\mu$ does not contribute in the coincidence limit, we may employ $\sigma_\mu$ as an expansion parameter. Coming from the transverse vector projectors, we have four covariant derivatives acting on this expression. Hence, an expansion up to order four in $\sigma_\mu$ is needed. Expanding the structure function as
\begin{equation}\label{VectorExpansionCoefficients}
	S^T (\sigma) = \sum_{n = 0}^\infty S^T_n \sigma^n \,,
\end{equation}
the expansion of $\mathcal{R}_{\alpha \beta'} G_{\chi_1, \chi_2} (\sigma) $ gives
\begin{equation}\label{ExpansionVector}
	\begin{split}
		\mathcal{R}_{\alpha \beta'} G_{\chi_1, \chi_2} (\sigma) =& S^T_0 \left( g_{\alpha \beta'} + \frac{R}{2 d (d - 1)} \sigma_\alpha \sigma_{\beta'} - \frac{\sigma R^2}{12 d^2 (d - 1)^2} \sigma_\alpha \sigma_{\beta'} \right) \\
		& + S^T_1 \left( \sigma g_{\alpha \beta'} + \frac{\sigma R}{2 d (d - 1)} \sigma_\alpha \sigma_{\beta'} \right) + S^T_2 \sigma^2 g_{\alpha \beta'} + \mathcal{O} \left( \sigma_\mu \right)^5 \, .
	\end{split}
\end{equation}
After acting with the projectors and contracting with $g^{\mu \nu'}$, we find
\begin{equation}\label{ExpansionVector2}
	g^{\mu \nu'} G_{Q, \mu \nu'}^T (0) = \frac{8 - 5d}{3} R\, S^T_1 + 2 d (d^2 + d - 2)\, S^T_2 \, .
\end{equation}
Explicit expressions for the  expansion coefficients $S^T_1$ and $S^T_2$ can be found in App.~\ref{sec:vecstrucfunc}. Inserting the expressions for $S^T_1$ and $S^T_2$ and using \eq{Gammasimplify}, we find
\begin{equation}\label{vectorkernel}
	\begin{split}
		&g^{\mu \nu'} G_{Q, \mu \nu'}^T (0) \\
		 =& 
		 \frac{1}{(4\pi)^{d/2}}
		  \left(\frac{d (d-1)}{R}\right)^{1-d/2} \left[ \frac{R \,\Gamma (d-1)}{\Gamma \left(\frac{d}{2}\right) (d Q-R)} 
		+\frac{\pi \left(d^2 Q-d Q-R\right)}{\sin \left(\frac{\pi  d}{2}\right) \Gamma \left(\frac{d}{2}\right) (d Q-R)} \sum _{k=0}^{\infty } \kappa _k(d)
   \left(-\overline{\xi}^2\right)^{d/2- k-1} \right] \,,
	\end{split}
\end{equation}
with
\begin{equation}\label{xibar}
	\overline{\xi} = \frac{1}{2} \sqrt{d \left(-\frac{4 (d-1) Q}{R}+d-2\right)+5} \, .
\end{equation}
To expand \eq{vectorkernel} in the large $m$ limit, we use the same trick as in the scalar case and set  $Q = m^2 - E$ for an  endomorphism such that $\overline{\xi}$ is directly proportional to $m$. We find
\beq\label{EbarV}
\begin{array}{rcl}
		\overline{E} &=&\displaystyle \frac{-5  + 2 d  - d^2 }{4 d (d - 1)}R \, , \\[2ex]
		-\overline{\xi}^2&=&\displaystyle d (d - 1) \frac{m^2}{R} \, .
	\end{array}
\end{equation}
Using \eq{EbarV} and the geometric series to expand denominators containing $m$ in the large $m$ limit, we arrive at
\begin{equation}\label{FinalVectorExpansion}
	\begin{split}
		g^{\mu \nu'} G^T_{Q, \mu \nu'} (0) \Big|_{\overline{E}}
		=& \frac{(d - 1)^2}{(4 \pi)^{d/2}} \frac{\pi}{\sin(d \pi/2) \Gamma(d/2)} \left( \frac{1}{d - 1} + \frac{R}{4 d m^2} \right) \times \\
		& \left( m^2 \right)^{d/2 - 1} \sum_{k = 0}^\infty \left( \frac{R}{m^2} \right)^{k} \sum_{\ell = 0}^k \left( - \frac{(d - 3)^2}{4} \right)^\ell \frac{\kappa_{k - \ell} (d)}{(d (d - 1))^k} \\
		& + \frac{R \, \Gamma(d - 1)}{d m^2 (4 \pi)^{d/2}} \frac{\left( \frac{d (d - 1)}{R} \right)^{1 - d/2}}{\Gamma(d/2)} \sum_{k = 0}^\infty \left( - \frac{(d - 3)^2}{4 d (d - 1)} \frac{R}{m^2} \right)^k \, ,
	\end{split}
\end{equation}
where we have reorganised sums coming from geometric series and \eq{Gammasimplify} to combine powers of $m$. As before, the transverse vector heat kernels $b_{2 n}^{(1)}$ and $c_{d + 2n}^{(1)}$ for the  endomorphism \eq{EbarV} can now be read off from  \eq{FinalVectorExpansion} by noticing that the coefficients $b_{2n}^{(1)}$ ($c_{d + 2n}^{(1)}$) are linear in  (independent of) the parameters $\kappa_i(d)$. We find
\begin{equation}
	\begin{split}
		b_{2 n}^{(1)}(\overline{E})
		=& \frac{(d - 1) \pi\,R^n }{\Gamma(1 + n - d/2)} \Bigg[ \sum_{\ell = 0}^n \left( - \frac{(d - 3)^2}{4} \right)^\ell \frac{\kappa_{n - \ell} (d) (d (d - 1))^{- n}}{\sin(d \pi/2) \Gamma(d/2)} \\
		& \qquad + \frac{(d - 1)^2}{4} \sum_{\ell = 0}^{n - 1} \left( - \frac{(d - 3)^2}{4} \right)^\ell \frac{\kappa_{n - 1 -\ell} (d) (d (d - 1))^{- n}}{\sin(d \pi/2) \Gamma(d/2)} \Bigg] \, , \\
		c_{d + 2n}^{(1)} (\overline{E})
		=& \frac{\Gamma(d)}{\Gamma(d/2) \, \Gamma(1 + n)} \left( \frac{R}{d (d - 1)} \right)^{d/2} \left( - \frac{(d - 3)^2}{4 d (d - 1)} R \right)^n \, ,
	\end{split}
	\label{eqn:resvector}
\end{equation}
and $\kappa_n (d)$ determined through \eq{kappa}. Applying \eq{eqn:heatkgeneralendo}, we then find the transverse vector heat kernel coefficients for general endomorphisms as well. Tab.~\ref{tab:vectorHK} shows our results for the first few vector heat kernels ($E=0$) and for a selection of dimensions.

\begin{table}
	\addtolength{\tabcolsep}{8pt}
	\setlength{\extrarowheight}{8pt}
	\begin{tabular}{`c?c|c|c|c|c`}
		\toprule
		\rowcolor{Yellow} & $\bm{d = 2}$ & $\bm{d = 3}$ & $\bm{d = 4}$ & $\bm{d = 5}$ & $\bm{d = 6}$  \\[1ex] \midrule
		$\bm{b_{0}^{(1)}}$ & $1$ & $2$ & $3$ & $4$ & $5$ \\[1ex]
		\rowcolor{LightGray} $\bm{c_{d}^{(1)}}$ & $\frac{1}{2}R$ & $\frac{2}{3\sqrt{6 \pi }} R^{3/2}$ & $\frac{1}{24}R^2$ & $\frac{1}{25 \sqrt{5 \pi }}R^{5/2}$ & $\frac{1}{450}R^3$ \\[1ex]
		$\bm{b_{2}^{(1)}}$ & $-\frac{1}{3}R$ & $0$ & $\frac{1}{4}R$ & $\frac{7}{15}R$ & $\frac{2}{3}R$ \\[1ex]
		\rowcolor{LightGray}$\bm{c_{d + 2}^{(1)}}$ & $\frac{1}{4}R^2$ & $\frac{2}{9\sqrt{6 \pi }} R^{5/2}$ & $\frac{1}{96}R^3$ & $\frac{1}{125 \sqrt{5 \pi }}R^{7/2}$ & $\frac{1}{2700}R^4$ \\[1ex] 
		$\bm{b_{4}^{(1)}}$ & $-\frac{11}{40}R^2$ & $-\frac{1}{9}R^2$ & $-\frac{67}{1440}R^2$ & $-\frac{1}{120}R^2$ & $\frac{7}{360}R^2$ \\[1ex] 
		\rowcolor{LightGray}$\bm{c_{d + 4}^{(1)}}$ & $\frac{1}{16}R^3$ & $\frac{1}{27 \sqrt{6 \pi }}R^{7/2}$ & $\frac{1}{768}R^4$ & $\frac{1}{1250 \sqrt{5 \pi }}R^{9/2}$ & $\frac{1}{32400}R^5$ \\[1ex] 
		$\bm{b_{6}^{(1)}}$ & $-\frac{37}{504}R^3$ & $-\frac{2}{81}R^3$ & $-\frac{4321}{362880}R^3$ & $-\frac{1}{160}R^3$ & $-\frac{2539}{850500}R^3$ \\[1ex] 
		\rowcolor{LightGray}$\bm{c_{d + 6}^{(1)}}$ & $\frac{1}{96}R^4$ & $\frac{1}{243 \sqrt{6 \pi }}R^{9/2}$ & $\frac{1}{9216}R^5$ & $\frac{1}{18750 \sqrt{5 \pi }}R^{11/2}$ & $\frac{1}{583200}R^6$ \\[1ex] 
		$\bm{b_{8}^{(1)}}$ & $-\frac{157}{13440}R^4$ & $-\frac{1}{324}R^4$ & $-\frac{3397}{2488320}R^4$ & $-\frac{17}{23040}R^4$ & $-\frac{176107}{408240000}R^4$ \\[1ex] 
		\rowcolor{LightGray}$\bm{c_{d + 8}^{(1)}}$ & $\frac{1}{768}R^5$ & $\frac{1}{2916 \sqrt{6 \pi }}R^{11/2}$ & $\frac{1}{147456}R^6$ & $\frac{1}{375000 \sqrt{5 \pi }}R^{13/2}$ & $\frac{1}{13996800}R^7$ \\[1ex] 
		$\bm{b_{10}^{(1)}}$ & $-\frac{109}{88704}R^5$ & $-\frac{1}{3645}R^5$ & $-\frac{245461}{2299207680}R^5$ & $-\frac{1}{18432}R^5$ & $-\frac{213749}{6735960000}R^5$ \\[1ex] 
		\rowcolor{LightGray}$\bm{c_{d + 10}^{(1)}}$ & $\frac{1}{7680}R^6$ & $\frac{1}{43740 \sqrt{6 \pi }}R^{13/2}$ & $\frac{1}{2949120}R^7$ & $\frac{1}{9375000 \sqrt{5 \pi }}R^{15/2}$ & $\frac{1}{419904000}R^8$ \\[1ex]
		\bottomrule
	\end{tabular}
	\caption{The vector heat kernel coefficients for different integer dimensions and vanishing endomorphism.}
	\label{tab:vectorHK}
\end{table}

\subsection{Transverse Traceless Tensors}\label{sec:restensor}
Turning to  the heat kernel coefficients for transverse traceless symmetric tensors, we first need to derive the Green's function. Just as in the case of transverse vectors, it is important to notice that the differential equation for the corresponding Green's function $G_{Q, \mu \nu \rho' \sigma'}^{TT}$ takes the form
\begin{equation}
	(- \nabla^2 + Q) G_{Q, \mu \nu \rho' \sigma'}^{TT} = \frac{g_{\mu ( \rho'} g_{\sigma') \nu}}{\sqrt{g}} \delta (x - y) + \text{longitudinal and trace terms} \, ,
	\label{eqn:tensordgl}
\end{equation}
where longitudinal terms include terms which are longitudinal with respect to at least one index but not necessarily all indices. Following the methods introduced previously
we write the transverse traceless symmetric Green's function as
\begin{equation}\label{GTT}
	G_{Q, \mu \nu \rho' \sigma'}^{TT} = \mathcal{P}_{\mu \nu}^{\,\,\,\,\,\,\alpha \beta} \mathcal{P}_{\rho' \sigma'}^{\,\,\,\,\,\,\,\,\gamma' \delta'} \left( \mathcal{R}_{\alpha \gamma'} \mathcal{R}_{\beta \delta'} S^{TT} \right) \, .
\end{equation}
The transverse traceless tensor projector is given by \cite{Mora:2012zi,Miao:2011fc}
\begin{equation}
	\begin{split}
		\mathcal{P}_{\mu \nu}^{\,\,\,\,\,\,\alpha \beta} =& \frac{1}{2} \frac{d - 3}{d - 2} \Bigg[ - g^\alpha_{(\mu} g^\beta_{\nu)} \left( \nabla^2 - \frac{R}{d - 1} \right) \left( \nabla^2 - \frac{2 R}{d (d - 1)} \right) \\
		& \qquad + \frac{g_{\mu \nu} g^{\alpha \beta}}{d - 1} \left( \nabla^4 - \frac{R}{d} \nabla^2  + \frac{2 R^2}{d^2 (d - 1)} \right) - \frac{g^{\alpha \beta}}{d - 1} \nabla_{(\mu} \nabla_{\nu)} \left( \nabla^2 + \frac{2 R}{d} \right) \\
		& \qquad + 2 \nabla_{(\mu} \left( \nabla^2 + \frac{R}{d (d - 1)} \right) g_{\nu )}^{(\alpha} \nabla^{\beta)} - \frac{d - 2}{d - 1} \nabla_{(\mu} \nabla_{\nu)} \nabla^{(\alpha} \nabla^{\beta)} \\
		& \qquad - \frac{g_{\mu \nu}}{d - 1} \left( \nabla^2 + \frac{2 R}{d} \right) \nabla^{(\alpha} \nabla^{\beta)} \Bigg] \, ,
	\end{split}
\end{equation}
and it satisfies
\begin{equation}
	\begin{gathered}
		\mathcal{P}_{\mu \nu}^{\,\,\,\,\,\,\alpha \beta} g_{\alpha \beta} S (\sigma) = g^{\mu \nu} \mathcal{P}_{\mu \nu}^{\,\,\,\,\,\,\alpha \beta} T_{\alpha \beta} = 0 \, , \qquad [\nabla^2, \mathcal{P}_{\mu \nu}^{\,\,\,\,\,\,\alpha \beta}] T_{\alpha \beta} = 0 \, , \\
		\mathcal{P}_{\mu \nu}^{\,\,\,\,\,\,\alpha \beta} \nabla_\alpha T_\beta = \mathcal{P}_{\mu \nu}^{\,\,\,\,\,\,\alpha \beta} \nabla_\beta T_\alpha = \nabla^\mu \mathcal{P}_{\mu \nu}^{\,\,\,\,\,\,\alpha \beta} T_{\alpha \beta} = \nabla^\nu \mathcal{P}_{\mu \nu}^{\,\,\,\,\,\,\alpha \beta} T_{\alpha \beta} = 0 \, , \\
		\mathcal{P}_{\mu \nu}^{\,\,\,\,\,\,\rho \sigma} \mathcal{P}_{\rho \sigma}^{\,\,\,\,\,\,\alpha \beta} T_{\alpha \beta} = - \frac{1}{2} \frac{d - 3}{d - 2} \mathcal{P}_{\mu \nu}^{\,\,\,\,\,\,\alpha \beta} \left( \nabla^2 - \frac{2 R}{d (d - 1)} \right) \left( \nabla^2 - \frac{R}{d - 1} \right) T_{\alpha \beta}
	\end{gathered}
\end{equation}
for an arbitrary tensor $T_{\alpha \beta}$. The Green's function can now be determined in a very similar fashion to the transverse vector Green's function. Firstly, we contract \eq{eqn:tensordgl} with $\mathcal{P}_{\lambda \eta}^{\,\,\,\,\,\, \mu \nu} \mathcal{P}_{\kappa' \chi'}^{\,\,\,\,\,\, \rho' \sigma'}$. Making use of the relations for the transverse traceless tensor projector, the longitudinal and trace terms on the right-hand side vanish and we get
\begin{equation}
	\begin{split}
		& \mathcal{P}_{\lambda \eta}^{\,\,\,\,\,\, \mu \nu} \mathcal{P}_{\kappa' \chi'}^{\,\,\,\,\,\, \rho' \sigma'} \bigg(- \nabla^2 + Q \bigg) \left( \nabla^2 - \frac{2 R}{d (d - 1)} \right)^2 \left( \nabla^2 - \frac{R}{d - 1} \right)^2 \bigg( \mathcal{R}_{\mu \rho'} \mathcal{R}_{\nu \sigma'} S^{TT} (\sigma) \bigg) \\
		=& 4 \left( \frac{d - 2}{d - 3} \right)^2 \mathcal{P}_{\lambda \eta}^{\,\,\,\,\,\, \mu \nu} \mathcal{P}_{\kappa' \chi'}^{\,\,\,\,\,\, \rho' \sigma'} \bigg( \mathcal{R}_{\mu \rho'} \mathcal{R}_{\nu \sigma'} \frac{1}{\sqrt{g}} \delta (x - y) \bigg) \, .
	\end{split}
\end{equation}
Using \eq{eqn:vectorprojectorprop}, this can be brought into the form
\begin{equation}
	\begin{split}
		& \mathcal{P}_{\lambda \eta}^{\,\,\,\,\,\, \mu \nu} \mathcal{P}_{\kappa' \chi'}^{\,\,\,\,\,\, \rho' \sigma'} \left[ \mathcal{R}_{\mu \rho'} \mathcal{R}_{\nu \sigma'} \bigg(- \nabla^2 + Q - \frac{2 R}{d (d - 1)} \bigg) \nabla^4\left( \nabla^2 - \frac{(d - 2)R}{d(d - 1)} \right)^2 S^{TT} (\sigma) \right] \\
		=& 4 \left( \frac{d - 2}{d - 3} \right)^2 \mathcal{P}_{\lambda \eta}^{\,\,\,\,\,\, \mu \nu} \mathcal{P}_{\kappa' \chi'}^{\,\,\,\,\,\, \rho' \sigma'} \bigg[ \mathcal{R}_{\mu \rho'} \mathcal{R}_{\nu \sigma'} \frac{1}{\sqrt{g}} \delta (x - y) \bigg] \,,
	\end{split}
\end{equation}
from which we deduce that $S^{TT} (\sigma) $ obeys the differential equation
\begin{equation}\label{GreenTensor}
	\bigg(- \nabla^2 + Q - \frac{2 R}{d (d - 1)} \bigg) \nabla^4\left( \nabla^2 - \frac{(d - 2)R}{d(d - 1)} \right)^2 S^{TT} (\sigma) = \frac{4}{\sqrt{g}} \left( \frac{d - 2}{d - 3} \right)^2 \delta (x - y) \, .
\end{equation}
The differential equation is solved by
\begin{equation}\label{SolutionTensor}
		S^{TT} (\sigma) = 4 \left( \frac{d - 2}{d - 3} \right)^2 G_{\chi_1, \chi_2, \chi_2, \chi_3, \chi_3} (\sigma)
		\eeq
		where we have introduced
		\beq
		\begin{array}{rcl}
		G_{\chi_1, \chi_2, \chi_2, \chi_3, \chi_3} (\sigma) &=& \frac{\partial}{\partial \chi_2} \frac{\partial}{\partial \chi_3} G_{\chi_1, \chi_2, \chi_3} (\sigma) \, , \\[2ex]
		G_{\chi_1, \chi_2, \chi_3} (\sigma) &=& \frac{G_{\chi_1, \chi_3} (\sigma) - G_{\chi_1, \chi_2} (\sigma) }{\chi_2 - \chi_3}\,, 
	\end{array}
\end{equation}
together with \eq{eqn:scalarGreen} and \eq{Gchi1chi2}, and evaluated for 		
\beq
		\chi_1 = Q - \frac{2R}{d (d - 1)}\,, \quad 
		\chi_2 =0 \,, \quad 
		\chi_3 = \frac{d-2}{d (d - 1)} R \,.
	\label{eqn:yensorstructuresolve}
\eeq
To calculate the trace of $G_{Q, \mu \nu \rho' \sigma'}^{TT}$ in the coincidence limit, we first expand the term $\mathcal{R}_{\mu \rho'} \mathcal{R}_{\nu \sigma'} S^{TT} (\sigma)$ as before. Since we are now acting with eight covariant derivatives on this expression, see \eq{SolutionTensor}, we have to expand it up to order eight in $\sigma_\mu$. The relevant expression is  lengthy and given in App.~\ref{sec:vecstrucfunc}, see \eq{ExpansionTensor8}, where we also  introduce
\begin{equation}\label{TensorExpansionCoefficients}
	S^{TT} (\sigma) = \sum_{n = 0}^\infty S^{TT}_n \sigma^n \, .
\end{equation}
Acting with the transverse tensor projectors and contracting the result with $g^{\mu (\rho'} g^{\sigma') \nu}$, we obtain
\begin{equation}\label{TensorContract}
	\begin{split}
		g^{\mu (\rho'} g^{\sigma') \nu} G_{Q, \mu \nu \rho' \sigma'}^{TT} (0) = & \frac{(d-3)^2 (d+1) (d+2)}{d-2} \Bigg[ \frac{(13 d-32) R^3}{140 (d-1)^2 d^2} S^{TT}_1 +\frac{(d-6) (7 d-8) R^2}{20 (d-1)^2 d} S^{TT}_2  \\
		& \qquad + \frac{3 (3-2 d) (d+4) R}{2 (d-1)} S^{TT}_3 + 3 d (d+4) (d+6) S^{TT}_4 \Bigg] \, .
	\end{split}
\end{equation}
Inserting the expansion coefficients for $S^{TT} (\sigma)$ and using \eq{Gammasimplify}, we finally arrive at
\begin{equation}\label{TensorFinal}
	\begin{split}
		&
		g^{\mu (\rho'} g^{\sigma') \nu} G_{Q, \mu \nu \rho' \sigma'}^{TT} (0)
		 \\
		=& \frac{2^{-d-1} (d-1) \pi ^{-\s0d2}}{\Gamma\left(\frac{d}{2}\right) ((d-1) Q-R)} \left(\frac{R}{d (d-1)}\right)^{d/2} \times \\
		& \Bigg[\frac{\Gamma (d+2) \left((d-1) d (d+2) Q-\left(d^2+4\right) R\right)}{d ((d-1) d Q-2 R)} \\
		& \qquad + \frac{\pi  (d-2) (d+1) ((d-1) d Q+(d-2) R)}{\sin
   \left(\frac{\pi  d}{2}\right) R} 
   		\sum _{k=0}^{\infty } \kappa _k (d) \left(
   		-\widetilde{\xi}^2
   		\right)^{d/2-k-1}\Bigg] \, .
	\end{split}
\end{equation}
where
\beq
 \widetilde{\xi}= \frac{1}{2}\sqrt{ \left(-\frac{4 (d-1) Q}{R}+d-2\right)d + 9} 
\eeq
Just as before, we now choose the endomorphism in order to simplify the sum. Taking
\begin{equation}\label{EbarT}
	\begin{array}{rcl}
		\overline{E} &=& \displaystyle - \frac{9 - 2d + d^2}{4 d (d - 1)} R \, , \\[2ex]
		-\widetilde{\xi}^2 
		&=& \displaystyle d (d - 1) \frac{m^2}{R} \,,
		\end{array}
\end{equation}
and using \eq{EbarT} and the geometric series to expand the denominators in the large $m$ limit, we find
\begin{equation}\label{TensorExpansion}
	\begin{split}
		&g^{\mu (\rho'} g^{\sigma') \nu} G_{Q, \mu \nu \rho' \sigma'}^{TT} (0) \Bigg|_{E = \overline{E}}
		 \\
		=& \frac{2^{-d-2} \pi ^{-d/2}}{m^2 R \Gamma \left(\frac{d}{2}-1\right)} \left(\frac{R}{d (d-1)}\right)^{1 + d/2} \sum_{n = 0}^\infty \left( - \frac{(d - 3)^2}{4 d (d - 1)} \frac{R}{m^2} \right)^n \times \\
		&\Bigg[\frac{\pi  d \left(d^2-1\right) \left(4 (d-1) d m^2+(d+1)^2 R\right)}{\sin \left(\frac{\pi  d}{2}\right) R} \sum
   _{k=0}^{\infty } \kappa _k(d) \left(\frac{d (d-1) m^2}{R}\right)^{d/2-k-1} \\ 
		& \qquad +\frac{\Gamma (d+2) \left(4 (d-1) d (d+2) m^2+(d ((d-4) d+5)+2) R\right)}{(d-2) d m^2} \sum_{m = 0}^\infty \left( - \frac{d - 1}{4 d} \frac{R}{m^2} \right)^m \Bigg] \\
		=& \frac{1}{(4 \pi)^{d/2}} \left( \frac{R}{d (d - 1)} \right)^{d/2} \frac{1}{\Gamma(d/2 - 1) } \sum_{k = 0}^\infty \left( \frac{R}{m^2} \right)^{k} \times \\
		& \left[ \frac{\Gamma(d + 2)}{(d - 2)} \left( \frac{d + 2}{dm^2} + \frac{(2 + d(5 + d(d - 4))) R}{4 d^2 (d - 1)m^4} \right) \sum_{\ell = 0}^k \left( - \frac{(d - 3)^2}{4 d (d - 1)} \right)^\ell \left(  - \frac{d - 1}{4 d} \right)^{k - \ell} \right. \\
		& \qquad + \left. \frac{\pi (d (d - 1))^{-k}}{\sin(d \pi /2)} \left( \frac{(d + 1)^3}{4 m^2} + \frac{d (d^2 - 1)}{R} \right) \left( \frac{R}{d (d - 1) m^2} \right)^{1 - d/2} \sum_{\ell = 0}^k \kappa_\ell (d) \left( - \frac{(d - 3)^2}{4} \right)^{k - \ell} \right] \, .
	\end{split}
\end{equation}
Note that in the last step sums have been reorganised in order to combine powers of $m$. As before, using \eq{TensorExpansion} and  the specific endomorphism \eq{EbarT},  the heat kernel coefficients are found to be
\begin{equation}
	\begin{split}
		b_{2 n}^{(2)} (\overline{E})
		=& \frac{1}{\Gamma(1 + n - d/2)} \left( \frac{R}{d (d - 1)} \right)^n \frac{\pi (d + 1)}{\Gamma(d/2 - 1) \sin(d \pi /2)} \times \\
		& \left[ \sum_{\ell = 0}^n \kappa_\ell (d) \left( - \frac{(d - 3)^2}{4} \right)^{n - \ell} + \frac{(d + 1)^2}{4} \sum_{\ell = 0}^{n - 1} \kappa_\ell (d) \left( - \frac{(d - 3)^2}{4} \right)^{n - 1 - \ell} \right] \, , \\ \\
		c_{d + 2n}^{(2)} (\overline{E})
		=& \frac{2}{\Gamma(1 + n)} \left( \frac{R}{d (d - 1)} \right)^{d/2} \frac{\Gamma(d + 2)}{\Gamma(d/2)} R^n \times \\
		&  \left[ \frac{d + 2}{4 d}  \sum_{\ell = 0}^{n} \left( - \frac{(d - 3)^2}{4 d (d - 1)} \right)^\ell \left(  - \frac{d - 1}{4 d} \right)^{n - \ell} \right. \\
		& \qquad \left. + \frac{2 + d(5 + d(d - 4))}{16 d^2 (d - 1)} \sum_{\ell = 0}^{n - 1} \left( - \frac{(d - 3)^2}{4 d (d - 1)} \right)^\ell \left(  - \frac{d - 1}{4 d} \right)^{n - 1 - \ell} \right] \, ,
	\end{split}
	\label{eqn:restensor}
\end{equation}
with $\kappa_n(d)$ determined through \eq{kappa}. As has  been observed for the scalar and transverse vector heat kernel coefficients, the  coefficients $b^{(2)}_{2n}$ are identified from \eq{TensorExpansion}  as the terms linear  in  $\kappa_i (d)$ while the coefficients $c^{(2)}_{d + 2n}$ are found by setting all $\kappa_i (d)$ to zero. The heat kernel coefficients for arbitrary endomorphisms can be found using \eq{eqn:heatkgeneralendo}.
Tab.~\ref{tab:tensorHK} shows our results for the first few tensor heat kernels $(E=0)$ and for a selection of dimensions. 

As a final remark, we emphasize that the explicit expressions for the heat kernel coefficients of scalar  \eq{bcScalarBar}, transverse vectors \eq{eqn:resvector}, and transverse traceless tensors \eq{eqn:restensor}, 
 are  easy to evaluate and implement for arbitrary endomorphism and arbitrary dimension, including non-integer ones. For example, the first 100  heat kernel coefficients \eq{eqn:restensor} at vanishing endomorphism in $d=4$ are found within 20 sec on a 3 GHz thread, also using    \eq{eqn:heatkgeneralendo} and the coefficients \eq{kappa}.

\begin{table}
	\addtolength{\tabcolsep}{8pt}
	\setlength{\extrarowheight}{8pt}
	\begin{tabular}{`c?c|c|c|c|c`}
\toprule 
\rowcolor{Yellow} & $\bm{d = 2}$ & $\bm{d = 3}$ & $\bm{d = 4}$ & $\bm{d = 5}$ & $\bm{d = 6}$  \\[1ex] \midrule
$\bm{b_{0}^{(2)}}$ & $0$ & $2$ & $5$ & $9$ & $14$ \\[1ex] 
\rowcolor{LightGray} $\bm{c_{d}^{(2)}}$ & $3R$ & $\frac{20}{3\sqrt{6 \pi }} R^{3/2}$ & $\frac{5}{8}R^2$ & $\frac{21}{25 \sqrt{5 \pi }}R^{5/2}$ & $\frac{14}{225}R^3$ \\[1ex] 
$\bm{b_{2}^{(2)}}$ & $-3R$ & $-\frac{5}{3}R$ & $-\frac{5}{6}R$ & $0$ & $\frac{14}{15}R$ \\[1ex] 
\rowcolor{LightGray} $\bm{c_{d + 2}^{(2)}}$ & $3R^2$ & $\frac{8}{3\sqrt{6 \pi }} R^{5/2}$ & $\frac{5}{36}R^3$ & $\frac{3}{25 \sqrt{5 \pi }}R^{7/2}$ & $\frac{7}{1125}R^4$ \\[1ex] 
$\bm{b_{4}^{(2)}}$ & $-3R^2$ & $-\frac{13}{12}R^2$ & $-\frac{271}{432}R^2$ & $-\frac{81}{200}R^2$ & $-\frac{56}{225}R^2$ \\[1ex] 
\rowcolor{LightGray} $\bm{c_{d + 4}^{(2)}}$ & $\frac{3}{2}R^3$ & $\frac{5}{9 \sqrt{6 \pi }}R^{7/2}$ & $\frac{5}{288}R^4$ & $\frac{21}{2000 \sqrt{5 \pi }}R^{9/2}$ & $\frac{7}{16875}R^5$ \\[1ex] 
$\bm{b_{6}^{(2)}}$ & $-\frac{3}{2}R^3$ & $-\frac{7}{24}R^3$ & $-\frac{7249}{54432}R^3$ & $-\frac{81}{1000}R^3$ & $-\frac{3347}{60750}R^3$ \\[1ex] 
\rowcolor{LightGray} $\bm{c_{d + 6}^{(2)}}$ & $\frac{1}{2}R^4$ & $\frac{13}{162 \sqrt{6 \pi }}R^{9/2}$ & $\frac{25}{15552}R^5$ & $\frac{29}{40000 \sqrt{5 \pi }}R^{11/2}$ & $\frac{7}{303750}R^6$ \\[1ex] 
$\bm{b_{8}^{(2)}}$ & $-\frac{1}{2}R^4$ & $-\frac{29}{576}R^4$ & $-\frac{22571}{1306368}R^4$ & $-\frac{729}{80000}R^4$ & $-\frac{106429}{18225000}R^4$ \\[1ex] 
\rowcolor{LightGray} $\bm{c_{d + 8}^{(2)}}$ & $\frac{1}{8}R^5$ & $\frac{35}{3888 \sqrt{6 \pi }}R^{11/2}$ & $\frac{5}{41472}R^6$ & $\frac{133}{3200000 \sqrt{5 \pi }}R^{13/2}$ & $\frac{49}{45562500}R^7$ \\[1ex] 
$\bm{b_{10}^{(2)}}$ & $-\frac{1}{8}R^5$ & $-\frac{37}{5760}R^5$ & $-\frac{20117}{12317184}R^5$ & $-\frac{729}{1000000}R^5$ & $-\frac{1275757}{3007125000}R^5$ \\[1ex] 
\rowcolor{LightGray} $\bm{c_{d + 10}^{(2)}}$ & $\frac{1}{40}R^6$ & $\frac{97}{116640 \sqrt{6 \pi }}R^{13/2}$ & $\frac{17}{2239488}R^7$ & $\frac{641}{320000000 \sqrt{5 \pi }}R^{15/2}$ & $\frac{287}{6834375000}R^8$ \\[1ex] 
\bottomrule
	\end{tabular}
	\caption{The tensor heat kernel coefficients for different integer dimensions and vanishing endomorphism.}
	\label{tab:tensorHK}
\end{table}

\subsection{Unconstrained Fields}\label{sec:resunconstrained}
In the literature, heat kernel coefficients are often calculated for unconstrained fields. A relation between the heat kernels of constrained fields and unconstrained fields can be derived by decomposing unconstrained fields into constrained fields. 

For vectors, the unconstrained vector field $v_\mu$ can be decomposed into a transverse vector field $v_\mu^T$ and a longitudinal part $\nabla_\mu \eta$ through
\beq
	v_\mu = v_\mu^T + \nabla_\mu \eta \, .
\eeq
Note that the first mode of the scalar field $\eta$ is constant and does not contribute to $v_\mu$. Hence, this mode must be excluded later. Considering the Laplacian $-\nabla^2$ acting on $v_\mu$, we can write
\beq
	-\nabla^2 v_\mu = -\nabla^2 \left( v_\mu^T + \nabla_\mu \eta \right) = - \nabla^2 v_\mu^T + \nabla_\mu \left( - \nabla^2 - \frac{R}{d} \right) \eta
\eeq
on the sphere. It follows that the eigenvalue spectrum of $-\nabla^2$ acting on an unconstrained vector field is the sum of two parts. The first part is the eigenvalue spectrum of the Laplacian acting on a transverse vector field. The second part is the eigenvalue spectrum of the Laplacian acting on a scalar field shifted by $- R/d$ and the first mode being excluded. Thus,
\begin{equation}
	\text{Tr}_{V} e^{t \nabla^2} = \text{Tr}_1 e^{t \nabla^2} + \text{Tr}_0' e^{t (\nabla^2 + R/d)} \, ,
	\label{eqn:vectorunconconrelation}
\end{equation}
where primes at the trace denote the exclusion of lowest modes and $\text{Tr}_V$ denotes the trace of an unconstrained vector field. This allows the computation of heat kernel coefficients for unconstrained fields using our results. We get
\begin{equation}\label{Vunconstrained}
	\begin{split}
		\text{Tr}_{V} e^{t \nabla^2} =& \frac{\text{Vol}}{(4 \pi t)^{d/2}} \Bigg[ d +\frac{d}{6} R t + \frac{5 d^3-7 d^2+6 d-60}{360 (d-1) d} R^2 t^2 \\ 
		& \quad + \frac{35 d^5-112 d^4+187 d^3-1370 d^2+852 d-1008}{45360 (d-1)^2 d^2} R^3 t^3 \\
		& \quad + \frac{R^4 t^4}{5443200 (d-1)^3 d^3} \Big( 175 d^7-945 d^6+2389 d^5-15711 d^4+23464 d^3 \\
		& \qquad -35436 d^2+59760 d-62640 \Big) \\
		&\quad+ \frac{R^5 t^5}{359251200 (d - 1)^4 d^4} \Big(385 d^9-3080 d^8 +10714 d^7-68156 d^6+168793 d^5 \\
		& \qquad -308084 d^4+858996 d^3-958944 d^2+857232 d-798336 \Big)
		+ \mathcal{O} (R t)^6\Bigg] \, .
	\end{split}
\end{equation}
As anticipated, we observe from the result that all coefficients $\tilde c^{(1)}_{d+2n}$ and $c^{(1)}_{d+2n}$ vanish for unconstrained fields, see \eq{eqn:heatkernel} and \eq{translate}.

Turning to tensors and using the York decomposition, we decompose an unconstrained symmetric tensor field $T_{\mu \nu}$ into
\beq
	T_{\mu \nu} = T_{\mu \nu}^{TT} + \nabla_\mu \xi_\nu + \nabla_\nu \xi_\mu + \left( \nabla_\mu \nabla_\nu - \frac{1}{d} g_{\mu \nu} \nabla^2 \right) \sigma + \frac{1}{d} g_{\mu \nu} \eta \, ,
\eeq
with $T^{TT}_{\mu \nu}$ being a transverse traceless symmetric tensor, $\xi_\mu$ being a transverse vector, and $\sigma$ and $\eta$ being scalars. As in the case of vector fields, not all modes of this decomposition contribute to $T_{\mu \nu}^{TT}$. These modes which have to be excluded are (i) the $d (d + 1)/2$ Killing vectors of $\xi_\mu$ which satisfy $\nabla_\mu \xi_\nu + \nabla_\nu \xi_\mu = 0$ and originate from the lowest transverse vector modes, (ii) the constant (lowest) mode of $\sigma$, and (iii) the $d + 1$ scalars of the second lowest modes of $\sigma$ (see \tab{tab:eigvmulti} for the multiplicities of these modes) corresponding to the proper conformal Killing vectors $\zeta_\mu=\nabla_\mu\sigma$, which satisfy  $\nabla_\mu \zeta_\nu + \nabla_\nu \zeta_\mu -\s02d  g_{\mu\nu}\,\nabla^\tau \zeta_\tau= 0$ \cite{Mottola:1995sj} (see also \cite{Lauscher:2001ya}). Acting with the Laplacian on $T_{\mu \nu}$, we use
\beq
	\begin{split}
		\nabla^2 \left( \nabla_\mu \xi_\nu + \nabla_\nu \xi_\mu \right) =& \, 2 \, \nabla_{(\mu} \left( - \nabla^2 - \frac{d + 1}{d (d - 1)} R \right) \xi_{\nu)} \\
		- \nabla^2 \left( \nabla_\mu \nabla_\nu - \frac{1}{d} g_{\mu \nu} \nabla^2 \right) \sigma =& \left( \nabla_\mu \nabla_\nu - \frac{1}{d} g_{\mu \nu} \nabla^2 \right) \left( - \nabla^2 - \frac{2}{d - 1} R \right) \sigma \, ,
	\end{split}
\eeq
to arrive at
\beq
	\begin{split}
	- \nabla^2 T_{\mu \nu} =& - \nabla^2 T_{\mu \nu}^{TT} + 2 \, \nabla_{(\mu} \left( - \nabla^2 - \frac{d + 1}{d (d - 1)} R \right) \xi_{\nu)} \\
	& + \left( \nabla_\mu \nabla_\nu - \frac{1}{d} g_{\mu \nu} \nabla^2 \right) \left( - \nabla^2 - \frac{2}{d - 1} R \right) \sigma - \frac{1}{d} g_{\mu \nu} \nabla^2 \eta \, .
	\end{split}
\eeq
This implies
\begin{equation}
	\text{Tr}_T e^{t \nabla^2} = \text{Tr}_{2} e^{t \nabla^2} + \text{Tr}_{1}' e^{t \left(\nabla^2 + \frac{d + 1}{d (d - 1)} R \right)} + \text{Tr}_{0}'' e^{t \left( \nabla^2 + \frac{2}{d - 1} R \right)} + \text{Tr}_{0} e^{t \nabla^2} \, ,
	\label{eqn:unconcontensorrelation}
\end{equation}
where $\text{Tr}_T$ denotes the trace of the Laplacian w.r.t.~an unconstrained symmetric tensor field and two primes denote the exclusion of the two lowest modes. Using this and our results for the heat kernel coefficients of constrained fields we find
\begin{equation}\label{Tunconstrained}
	\begin{split}
		\text{Tr}_T e^{t \nabla^2} =& \frac{\text{Vol}}{(4 \pi t)^{d/2}} \Bigg[ \frac{d (d + 1)}{2} + \frac{d (d + 1)}{12} R t + \frac{5 d^4-2 d^3-d^2-114 d-240}{720 (d-1) d} R^2 t^2 \\
		& \quad + \frac{35 d^6-77 d^5+75 d^4-2443 d^3-3542 d^2+1104 d-4032}{90720 (d-1)^2 d^2} R^3 t^3 \\
		& \quad + \frac{R^4 t^4}{10886400 (d-1)^3 d^3} \Big( 175 d^8-770 d^7+1444 d^6-25922 d^5-9887 d^4 \\
		& \qquad +13588 d^3+188844 d^2+742320 d+172800\Big) \\
		& \quad + \frac{R^5 t^5}{718502400 (d-1)^4 d^4} \Big(385 d^{10}-2695 d^9+7634 d^8-103642 d^7+100637 d^6-8875 d^5 \\
		& \qquad +2850880 d^4+8146092 d^3+7406448 d^2+18339840 d+8211456\Big) 
		+ \mathcal{O} (R t)^6 \Bigg] \, .
	\end{split}
\end{equation}
Note  that all coefficients  $\tilde c^{(2)}_{d+2n}$ and $c^{(2)}_{d+2n}$ vanish for unconstrained fields.
This completes the derivation of heat kernel coefficients.

\section{\bf Heat Kernels from Spectral Sums}\label{sec:resspectralsum}
In this section, we compute heat kernel coefficients with the help of spectral sums and the Euler-Maclaurin formula. This serves as an independent consistency check for  findings in the previous section. We also find new expressions for certain heat kernel coefficients in the form of spectral integrals.

\subsection{Spectral Sum Technique}\label{SpectralSum}

\begin{table}
	\addtolength{\tabcolsep}{12pt}
	\setlength{\extrarowheight}{6pt}
	\begin{tabular}{`c?c|c`}
		\toprule
		\rowcolor{Yellow} \bf Spin $\bm s$ & \bf Eigenvalue $\bm{\lambda_\ell^s}$ & \bf Multiplicity $\bm{D_\ell^s}$ \\[1ex] \midrule
		$0$ & $ \frac{\ell (\ell + d - 1)}{d (d - 1)} R$;\quad  $\ell = 0, 1, 2, ...$ & $ \frac{(2\ell + d - 1) (\ell + d - 2)!}{\ell! (d - 1)!}$ \\[1ex]
		\rowcolor{LightGray} $1$ & $ \frac{\ell (\ell + d - 1) - 1}{d (d - 1)} R$; $\ell = 1, 2, 3, ...$ & $ \frac{\ell (\ell + d - 1) (2\ell + d - 1) (\ell + d - 3)!}{(d - 2)! (\ell + 1)!}$ \\ [1ex]
		$2$ & $ \frac{\ell (\ell + d - 1) - 2}{d (d - 1)} R$; $\ell = 2, 3, 4, ...$ & $ \frac{(d + 1) (d - 2) (\ell + d) (\ell - 1) (2 \ell + d - 1) (\ell + d - 3)!}{2 (d - 1)! (\ell + 1)!}$ \\ 
		\bottomrule 
	\end{tabular}
	\caption{The eigenvalues $\lambda_\ell^s$ and their multiplicities $D_\ell^s$ for spin $0$ (scalar), spin $1$ (transverse vector), and spin $2$ (transverse traceless tensor) fields of the operator $- \nabla^2$ on the sphere (taken from \cite{Rubin:1983be,Rubin:1984tc}).}
	\label{tab:eigvmulti}
\end{table}

The eigenspectrum of the Laplacian $- \nabla^2$ is known on the sphere  \cite{Rubin:1983be,Rubin:1984tc} and can be used to calculate the heat kernel coefficients, $e.g.$~\cite{Percacci:2017fkn,Benedetti:2014gja,Demmel:2014sga}. Specifically, for different spins $s$, the eigenfunctions $\phi_\ell^s$ satisfy the eigenvalue equation
\beq
	- \nabla^2\phi_\ell^s (x) = \lambda_\ell^s \, \phi_\ell^s (x) \, , 
\eeq
where $\lambda_\ell^s$ are the eigenvalues and all vector and tensor indices have been suppressed. In Tab.~\ref{tab:eigvmulti} we show the eigenvalues and their multiplicities $D_\ell^s$ for spin $0$, spin $1$, and spin $2$ fields on the sphere. The eigenfunctions can be chosen to be orthonormal,
\beq
	\int \text{d}^d x \, \sqrt{g} \, \phi_{\ell, n}^{s \, *} (x) \phi_{k, m}^s (x) = \delta_{\ell, k} \delta_{n, m} \, ,
\eeq
where we use the indices $n$ and $m$ to distinguish eigenfunctions with equal eigenvalues. Further, they satisfy the completeness relation
\beq
	\sum_{\ell, n} \phi_{\ell, n}^{s \, *} (x) \phi_{\ell, n}^s (y) = \frac{\delta (x - y)}{\sqrt{g}} \, .
\eeq
With this at hand, and for vanishing endomorphisms, we may  express the trace of the heat kernel as a multiplicity-weighted sum of its eigenvalues,
\beq
	\text{Tr}_s \, U_0 (t, \sigma) = \sum_{\ell, n} \int \text{d}^d x \, \sqrt{g} \, \phi_{\ell, n}^{s \, *} (x) e^{t \nabla^2} \phi_{\ell, n}^s (x) = \sum_\ell D_\ell^s \, e^{-t \lambda_\ell^s} \, .
	\label{eqn:HKeigvsum}
\eeq
In principle, the spectral sum  representation allows the extraction of  heat kernel coefficients by projection
\beq\label{b2nc2n}
\begin{split}
b^{(s)}_{2n}=&\left.\frac{(4\pi)^{d/2}}{\text{Vol}}\frac{1}{n!}\frac{\text{d}^n}{\text{d}t^n}\left(t^{d/2}\,\sum_\ell D_\ell^s \, e^{-t \lambda_\ell^s}\right)\right|_{t=0}\,,\\
c^{(s)}_{d+2n}=&\left.\frac{(4\pi)^{d/2}}{\text{Vol}}\frac{1}{n!}\frac{\text{d}^n}{\text{d}t^n}\left(\,\sum_\ell D_\ell^s \, e^{-t \lambda_\ell^s}\right)\right|_{t=0}\,.
	\end{split}
	\eeq
Note that in even dimension $d=2m$, the coefficients $b_{2(m+n)}$ and $c_{d+2n}$ cannot be distinguished. 
In practice, to find explicit expressions, the spectral sum in \eq{eqn:HKeigvsum} and \eq{b2nc2n}  is now approximated systematically using the Euler-Maclaurin formula 
\begin{equation}\label{Euler}
	\sum_{\ell = a}^b f(\ell) = \int_a^b \text{d} \ell \, f (\ell) + \frac{1}{2} \left( f (a) + f(b) \right) + \sum_{k = 1}^{n} \frac{B_{2k}}{(2k)!} \left( f^{(2k - 1)} (b) - f^{(2k - 1)} (a) \right) + R_{2n+1} \, ,
\end{equation}
where $B_n$ are the Bernoulli numbers (recall that $B_n=0$ for odd $n$ except $n=1$) and $f^{(n)} (\ell)\equiv \frac{\text{d}^n}{ \text{d}\ell^n}f( \ell)$;  see $e.g.$~\cite{Percacci:2017fkn} for a derivation of \eq{Euler}. An approximation of the Euler-Maclaurin formula at order $2n$ corresponds to dropping the remainder part 
\beq\label{remainder}
	R_{2n+1} = \frac{1}{(2n+1)!} \int_a^b \text{d} \ell \, f^{(2n+1)} (\ell) \, P_{2n+1}(\ell) \, ,
\eeq
where  $P_n(\ell)=B_n( \ell - \lfloor \ell \rfloor)$ are   periodized versions of the Bernoulli polynomials  $B_n(\ell)$. An upper bound for the remainder    is given by
\beq\label{rbound}
	|R_{2n+1}| \leq \frac{2}{(2\pi)^{2n}} \int_a^b \text{d} \ell \, |f^{(2n+1)} (\ell)| \,.
\eeq
In the  case at hand, we have $a=s$, $1/b=0$,  and $f(\ell) = D_\ell^s \, e^{-t \lambda_\ell^s}$, with eigenvalues and multiplicities as in Tab.~\ref{tab:eigvmulti}. 
The eigenvalues are quadratic polynomials in $\ell$  and imply  that the leading small~$t$ behaviour of the remainder   is controlled by the leading large~$\ell$ behaviour of the integrand, hence $\ell^2\sim t^{-1}$. In this limit,  multiplicities of eigenmodes for scalars, transverse vectors, and transverse traceless tensors  scale  according to  $D_\ell^s \sim \ell^{d-1}\sim t^{(1-d)/2}$.  Together with the elementary integral $\int^\infty_a \text{d}\ell \exp{-t \ell^2}=\sqrt{\frac{\pi}{4 t}}-a+{\cal O}(t\,a^3)$ we find a lower bound for the leading power in $t$ contained in the remainder,
\beq
	R_{2n+1}
	= \mathcal{O} \left( t^{ n+(1 - d)/2} \right) \,.
	\label{eqn:Rn_order}
\eeq
Therefore,  we may expect that truncating the Euler-Maclaurin formula \eq{Euler} to a finite order $2n$ and leaving out the remainder  $R_{2n+1}$ reproduces the series expansion in the proper time parameter for the heat kernel \eq{eqn:HKeigvsum} up to all orders including $ t^{(2n - d)/2}$. This corresponds to coefficients up to order $n$ in Ricci curvature $\sim \frac{\text{Vol}}{(4 \pi t)^{d/2}} (R\,t)^n$ in  the heat kernel expansion \eq{TrsU} with coefficients \eq{b2nc2n}.  Below, we also show that the bound is exhaustive. Finally, we note that the boundary and derivative terms  $f (a)$, $f (b)$, and  $ f^{(2k - 1)} (\ell)|^b_a$ only generate contributions proportional to integer positive powers in $t$, in any dimension.

\subsection{Results from Euler-Maclaurin}\label{SpectralSumResults}
The Euler-Maclaurin formula \eq{Euler} is most efficient in settings where the spectral integrals $\int_a^b \text{d} \ell \, f (\ell)$ can be solved in closed form. For our setting, this is the case in integer dimensions where the multiplicities are simple polynomials. 
The bound \eq{eqn:Rn_order} for the remainder   ensures that approximations  improve with increasing  order. Here, we  exploit these features  to cross-check results for heat kernel coefficients, order-by-order.  

Putting our rationale to work  for the scalar heat kernel coefficients in integer dimensions
by using the expression \eq{eqn:HKeigvsum} together with \eq{Euler}, we find
\beq\label{SSS}
	\begin{split}
		\text{Tr}_0 U_0 (t, \sigma) \bigg|_{d = 2} =& \, \frac{\text{Vol}}{4 \pi t} \left( 1+\frac{R t}{6}+\frac{R^2 t^2}{60}+\frac{R^3 t^3}{630}+\frac{R^4 t^4}{5040}+\frac{R^5 t^5}{27720}+\mathcal{O} (R t)^6 \right) \, , \\
		\text{Tr}_0 U_0 (t, \sigma) \bigg|_{d = 3} =& \, \frac{\text{Vol}}{(4 \pi t)^{3/2}} \left( 1+\frac{R t}{6}+\frac{R^2 t^2}{72}+\frac{R^3 t^3}{1296}+\frac{R^4 t^4}{31104}+\frac{R^5 t^5}{933120} + \mathcal{O}(R t)^6\right) \, , \\
		\text{Tr}_0 U_0 (t, \sigma) \bigg|_{d = 4} =& \, \frac{\text{Vol}}{(4 \pi t)^{2}} \left( 1+\frac{R t}{6}+\frac{29 R^2 t^2}{2160}+\frac{37 R^3 t^3}{54432}+\frac{149 R^4 t^4}{6531840}+\frac{179 R^5 t^5}{431101440}+\mathcal{O}(R t)^6 \right) \, , \\
		\text{Tr}_0 U_0 (t, \sigma) \bigg|_{d = 5} =& \, \frac{\text{Vol}}{(4 \pi t)^{5/2}} \left( 1+\frac{R t}{6}+\frac{R^2 t^2}{75}+\frac{R^3 t^3}{1500}+\frac{R^4 t^4}{45000}+\frac{R^5 t^5}{2250000}+\mathcal{O}(R t)^6 \right) \, , \\
		\text{Tr}_0 U_0 (t, \sigma) \bigg|_{d = 6} =& \, \frac{\text{Vol}}{(4 \pi t)^{3}} \left( 1+\frac{R t}{6}+\frac{R^2 t^2}{75}+\frac{1139 R^3 t^3}{1701000}+\frac{833 R^4 t^4}{36450000}+\frac{137 R^5 t^5}{267300000} + \mathcal{O} (R t)^6 \right) \,.
	\end{split}
\eeq
  To obtain results for heat kernel coefficients up to order $n=5$ in the Ricci curvature, we confirm that  an expansion of \eq{Euler} up to order $2n = 10$ is required, in agreement  with  the estimate for the remainder term  stated in \eq{eqn:Rn_order}. Also, results are in exact agreement with findings from the Green's function technique (see \tab{tab:scalarHK}) and confirm that the leading contribution starts out as $t^{-d/2}$.

The same analysis can now be done for transverse vectors and transverse traceless tensors. However, for these cases it is important to realise that in even dimensions we cannot distinguish between the contributions coming from the $b_{d+2n}^{(s)}$ and the $c_{d + 2n}^{(s)}$ $(n\ge 0)$ when we take the sum of the eigenvalues, see \eq{b2nc2n}. Thus, in even dimension we can only compare their sum with the findings from spectral sums. Again, to find  all coefficients $b_{2m}^{(s)}$ and the $c_{2m}^{(s)}$ up to $m=n$ we have to expand \eq{Euler} up to including the order $2n$, see \eq{eqn:Rn_order}.
For the transverse vectors in even dimension and for expansion order $2n=10$ we find using the spectral sum
\beq\label{SSTVe}
	\begin{split}
		\text{Tr}_1 U_0 (t, \sigma) \bigg|_{d = 2} =& \, \frac{\text{Vol}}{4 \pi t} \left( 1+\frac{R t}{6}-\frac{R^2 t^2}{40}-\frac{11 R^3 t^3}{1008}-\frac{17 R^4 t^4}{13440}+\frac{13 R^5 t^5}{177408} + \mathcal{O} (R t)^6 \right) \, ,\\
		\text{Tr}_1 U_0 (t, \sigma) \bigg|_{d = 4} =& \, \frac{\text{Vol}}{(4 \pi t)^{2}} \left( 3+\frac{R t}{4}-\frac{7 R^2 t^2}{1440}-\frac{541 R^3 t^3}{362880}-\frac{157 R^4 t^4}{2488320}+\frac{4019 R^5 t^5}{2299207680} + \mathcal{O} (R t)^6 \right) \, , \\
		\text{Tr}_1 U_0 (t, \sigma) \bigg|_{d = 6} =& \, \frac{\text{Vol}}{(4 \pi t)^{3}} \left( 5+\frac{2 R t}{3}+\frac{7 R^2 t^2}{360}-\frac{649 R^3 t^3}{850500}-\frac{24907 R^4 t^4}{408240000}-\frac{5849 R^5 t^5}{6735960000} + \mathcal{O} (R t)^6 \right) \, .
	\end{split}
\eeq
Taking the sum $b_{d+2n}^{(1)} + c_{d+2n}^{(1)}$ from \tab{tab:vectorHK}, the consistency of our results with the Geen's function method is confirmed.
In odd dimensions, we  can distinguish between the contributions of $b_{d+2n}^{(1)}$ and $c_{d + 2n}^{(1)}$ from within the spectral sum technique, and the results up to order five in Ricci curvature
\beq\label{SSTVo}
	\begin{split}
		\text{Tr}_1 U_0 (t, \sigma) \bigg|_{d = 3} =& \, \frac{\text{Vol}}{(4 \pi t)^{3/2}} \Bigg( 2+\frac{2 R^{3/2} t^{3/2}}{3 \sqrt{6 \pi }} -\frac{R^2 t^2}{9}+\frac{2 R^{5/2} t^{5/2}}{9 \sqrt{6 \pi }}-\frac{2}{81} R^3 t^3+\frac{R^{7/2} t^{7/2}}{27 \sqrt{6 \pi }} \\
		& \quad -\frac{1}{324} R^4 t^4+\frac{R^{9/2} t^{9/2}}{243 \sqrt{6 \pi }}-\frac{R^5 t^5}{3645} + \mathcal{O}(R t)^{11/2}\Bigg) \, , \\
		\text{Tr}_1 U_0 (t, \sigma) \bigg|_{d = 5} =& \, \frac{\text{Vol}}{(4 \pi t)^{5/2}} \Bigg( 4+\frac{7 R t}{15}-\frac{R^2 t^2}{120}+\frac{R^{5/2} t^{5/2}}{25 \sqrt{5 \pi }}-\frac{R^3 t^3}{160}+\frac{R^{7/2} t^{7/2}}{125 \sqrt{5 \pi }}-\frac{17 R^4 t^4}{23040} \\
		& \quad +\frac{R^{9/2} t^{9/2}}{1250 \sqrt{5 \pi }}-\frac{R^5 t^5}{18432} + \mathcal{O}(R t)^{11/2} \Bigg) \, .
	\end{split}
\eeq
fully agree with the corresponding results from the Green's function technique  (see \tab{tab:vectorHK}).
Finally, for the transverse traceless tensor heat kernels in even dimensions we get from the spectral sum approximation 
\beq\label{SSTTe}
	\begin{split}
		\text{Tr}_2 U_0 (t, \sigma) \bigg|_{d = 2} =& \, 0 \, ,\\
		\text{Tr}_2 U_0 (t, \sigma) \bigg|_{d = 4} =& \, \frac{\text{Vol}}{(4 \pi t)^{2}} \left( 5-\frac{5 R t}{6}-\frac{R^2 t^2}{432}+\frac{311 R^3 t^3}{54432}+\frac{109 R^4 t^4}{1306368}-\frac{317 R^5 t^5}{12317184} + \mathcal{O} (R t)^6 \right) \, , \\
		\text{Tr}_2 U_0 (t, \sigma) \bigg|_{d = 6} =& \, \frac{\text{Vol}}{(4 \pi t)^{3}} \left( 14+\frac{14 R t}{15}-\frac{56 R^2 t^2}{225}+\frac{433 R^3 t^3}{60750}+\frac{6971 R^4 t^4}{18225000}-\frac{28357 R^5 t^5}{3007125000} + \mathcal{O} (R t)^6 \right) \, .
	\end{split}
\eeq
Similar to the case of transverse vectors, the consistency of these results with \tab{tab:tensorHK} is confirmed by taking the sum $b_{d+2n}^{(2)} + c_{d+2n}^{(2)}$.
In odd dimensions we find
\beq\label{SSTTo}
	\begin{split}
		\text{Tr}_2 U_0 (t, \sigma) \bigg|_{d = 3} =& \, \frac{\text{Vol}}{(4 \pi t)^{3/2}} \Bigg( 2-\frac{5 R t}{3}+\frac{20}{3 \sqrt{6 \pi }} R^{3/2} t^{3/2}-\frac{13 R^2 t^2}{12}+\frac{8}{3\sqrt{6 \pi }} R^{5/2} t^{5/2}-\frac{7}{24} R^3 t^3 \\
		& \quad +\frac{5 R^{7/2} t^{7/2}}{9 \sqrt{6 \pi }}-\frac{29}{576} R^4 t^4+\frac{13 R^{9/2} t^{9/2}}{162 \sqrt{6 \pi }}-\frac{37 R^5 t^5}{5760} + \mathcal{O}(R t)^6\Bigg) \, , \\
		\text{Tr}_2 U_0 (t, \sigma) \bigg|_{d = 5} =& \, \frac{\text{Vol}}{(4 \pi t)^{5/2}} \Bigg( 9-\frac{81 R^2 t^2}{200}+\frac{21 R^{5/2} t^{5/2}}{25 \sqrt{5 \pi }}-\frac{81 R^3 t^3}{1000}+\frac{3 R^{7/2} t^{7/2}}{25 \sqrt{5 \pi }}-\frac{729 R^4 t^4}{80000} \\
		& \quad +\frac{21 R^{9/2} t^{9/2}}{2000 \sqrt{5 \pi }}-\frac{729 R^5 t^5}{1000000} + \mathcal{O}(R t)^6 \Bigg) \, ,
	\end{split}
\eeq
which is, again, in  agreement with \tab{tab:tensorHK}. This completes the derivation and checks of heat kernel coefficients on spheres in various integer dimensions.

\subsection{Spectral Integrals}\label{SpectralIntegrals}
We close with a few observations regarding  the usage of the Euler-Maclaurin formula   for spectral sums, and  spectral integrals for heat kernel coefficients in general dimensions. 
In integer dimensions, we find that all terms proportional to half-integer powers in $t$, if they arise,  originate from the integral term $ \int_a^b \text{d} \ell \, f (\ell) $ in the Euler-Maclaurin formula. This observation  implies that the remainder  $R_{2n+1}$ do not generate terms proportional to half-integer powers in  $t$.
 Moreover, in odd dimensions it also implies that  all heat kernel coefficients $b_{2n}$ are  determined from the expansion of the integral term only,
 offering the new representation 
\beq\label{b2nNew}
b^{(s)}_{2n}=\left.\frac{(4\pi)^{d/2}}{\text{Vol}}\frac{1}{n!}\frac{\text{d}^n}{\text{d}t^n}\left(t^{d/2}\int_s^\infty\text{d}\ell \,D^s_\ell\,e^{-t\,\lambda_\ell^s}\right)\right|_{t=0}
\eeq
instead of \eq{b2nc2n}. Note that the spectral sum has become a spectral integral. We rush to add that the expressions \eq{b2nNew} for $b_{2n}$ do not apply in even dimensions, and  that there is no analoguous formula for the coefficients $c_{d+2n}$, the simple reason being that these  coefficients receive contributions from both the integral term and from the boundary terms in the Euler-Maclaurin expansion. 

In odd dimensions, it is then straightforward to find  expressions for heat kernels by evaluating the spectral integral in closed form, followed by a projection onto those terms which in a small $t$ expansion generate half-integer powers in $t$. Following this strategy for scalar heat kernels  (which have no contributions from $c_{d+2n}$ terms),  exemplarily  in $d=3,5$ and 7 dimensions, we find 
\bea\label{d=3}
\sum_{n=0}^\infty b^{(0)}_{2n}(E)\, t^n \bigg|_{d = 3}&=&e^{(E+R/6)t}\,, \\
\label{d=5}
\sum_{n=0}^\infty b^{(0)}_{2n}(E)\, t^n \bigg|_{d = 5}&=&e^{(E+R/5)t} \left(1-\frac{R\, t}{30}\right)\,,\\ 
\label{d=7}
 \sum_{n=0}^\infty b^{(0)}_{2n}(E)\, t^n \bigg|_{d = 7}&=&e^{(E+\s03{14}R)t} \left(1-\frac{R\, t}{21}+\frac{4\, R^2\, t^2}{6615}\right)\,,
   \eea
 for arbitrary endomorphism $E$, and in  agreement  with \eq{eqn:resscalar}; see also Tab.~\ref{tab:scalarHK}. When multiplied with $\frac{\text{Vol}}{(4\pi\,t)^{d/2}}$, \eq{vol}, it provides closed expressions for the local part of the scalar heat kernels \eq{TrsU}. Similar closed expressions for the local heat kernels can be derived in any odd dimension using the spectral integral as discussed above. Note that on compact spaces such as $S^d$, the heat kernels also receives non-local "topological" contributions  \cite{Camporesi:1990wm}. These are not accounted for in the above.

\subsection{Analytic Continuation}
\label{sct:anacon}
In non-integer dimension, multiplicities are no longer  simple polynomials, and 
closed expressions for the $\ell$ integrals are not available.
Still, the coefficients $b_{2n}$ relate to non-integer powers in $t$ (in non-integer dimensions), and, once more, can only
arise from the integral term $ \int_a^b \text{d} \ell \, f (\ell) $. 
This suggests that \eq{b2nNew} can be extended to even dimensions by analytic continuation. 

We will now explain how  spectral integrals can be used in general  dimensions to find the heat kernel coefficients $b_{2n}$. The idea is to evaluate the integral initially in non-integer dimensions using a large $\ell$ expansion for the multiplicities in Tab.~\ref{tab:eigvmulti},  and interchanging sum and integration. The resulting finite expressions  can then be dimensionally continued to integer (even) dimension.\footnote{Since in all but even dimensions the coefficients $b_{2n}$ only arise from the spectral integral, it is not surprising that the coefficients $b_{2n}$ in even dimensions follow
 from dimensional continuation.} Our procedure  also eliminates all contributions to the coefficients $c_{2n}$. In other words, we have
\beq
\frac{\text{Vol}}{(4\pi\,t)^{d/2}}\sum_n b^{(s)}_{2n}\, t^n 
=\left.\int_0^\infty\text{d}\ell \,D^s_\ell\,e^{-t\,\lambda_\ell^s}\right|_{\rm reg.}\,,
\label{eqn:specint}
\eeq
where ``reg." indicates the procedure described above.
The integration starts at zero for any spin, the reason being that any finite  integral $\int_0^s{\rm d}\ell f(\ell)$ 
only generates terms in the form of integer powers of $t$ which contribute  to the coefficients $c_{2n}$. 
We apply the procedure  in App.~\ref{app:scalarspec}  for scalars in general dimensions including all details of the computation. 
The final result takes the form 
\beq
	\begin{split}
		\frac{\text{Vol}}{(4\pi\,t)^{d/2}}\sum_n b^{(0)}_{2n}\, t^n =& \,
		\0{1}{\Gamma(d)} \left( \0{d (d - 1)}{R\,t} \right)^{d/2}\times \\ & 
		\left\{\ \ \,
		\sum_{n = 0}^\infty 
		{d - 2 \choose 2n}
		B_{2n}^{(d - 1)} (d - 1) 
		\left[I_{d,2n}(R\,t)-\0{1}{2}(d-1)^2\,J_{d,2n+1}(R\,t)\right]\right.
		\\ & 
		\left. 
		+\0{d-1}2 \sum_{n = 0}^\infty
		 {d - 2 \choose 2n + 1}
		B_{2n + 1}^{(d - 1)} (d - 1) \bigg[I_{d,2n}(R\,t)-2J_{d,2n+1}(R\,t)\bigg]\right\}
	\end{split}
	\label{Spectral_b2n}
\eeq
where $B_n^{(\ell)}(x)$ are Bernoulli polynomials \eq{Bnlx}, and  the functions $I_{d,n}(x)$ and $J_{d,n}(x)$ relate to the Kummer hypergeometric series, \eq{IJ}. This result constitutes a  representation of heat kernel coefficients which is complementary  (yet, equivalent) to the result from the Green's function method \eq{eqn:resscalar}. A series expansion of \eq{Spectral_b2n} in $R$ yields all heat kernel coefficients $b_{2n}$  in general dimension, and agrees with all findings in integer dimensions reported earlier.  In the case where $d$ is an odd integer, the sums can be performed in closed form, and we find agreement with  the expressions \eq{d=3}, \eq{d=5}, and \eq{d=7}   in $d=3,5,7$ dimensions.

Finally, we emphasize once more that the
spectral integrals  with or without dimensional continuation are in general different,  $\int_0^\infty\text{d}\ell \,D^s_\ell\,e^{-t\,\lambda_\ell^s}\neq \left.\int_0^\infty\text{d}\ell \,D^s_\ell\,e^{-t\,\lambda_\ell^s}\right|_{\rm reg.}$. The reason for this is that, without dimensional continuation, the integration additionally generates contributions to the $c_{2n}$ coefficients, all of which are absent in 
 \eq{eqn:specint}. 
Further, in even dimensions the $b_{2n}$ coefficients also receive contributions from boundary terms and derivative terms in \eq{Euler}. In contrast to this, the analytical continuation from non-integer dimensions, giving rise to the right-hand side of \eq{eqn:specint}, contains all and only the coefficients $b_{2n}$ in any dimension. 

\section{\bf Discussion}\label{sec:Discussion}

With the help of Green's functions we have derived general expressions for all heat kernel coefficients of  scalars \eq{bcScalarBar}, transverse vectors \eq{eqn:resvector}, and transverse traceless tensors \eq{eqn:restensor} on the sphere in any dimension and for any endomorphism, also providing the corresponding results for unconstrained fields, see \eq{Vunconstrained}, \eq{Tunconstrained}.  The final expressions are easy to evaluate and straightforward to implement on a practical level, with explicit results stated for selected integer dimensions  (Tab.~\ref{tab:scalarHK},~\ref{tab:vectorHK} and~\ref{tab:tensorHK}). 
Several  consistency checks have  been performed. We compared the first five heat kernel coefficients and their  full dimensional dependence to the known results for heat kernel coefficients on general manifolds \cite{DeWitt:1965jb,Gilkey:1975iq,Christensen:1976vb,Christensen:1978yd,Amsterdamski:1989bt,Avramidi:1989ik,Avramidi:1990je,Avramidi:1990ug,vandeVen:1997pf}, and found complete agreement. 

Furthermore, we have   derived the local heat kernel coefficients on spheres from  
known eigenspectra of Laplacians. The boundedness of remainder  terms  \eq{eqn:Rn_order} in the  Euler-Maclaurin formula has been demonstrated, which is a prerequisite for its applicability to spectral sums. In  integer dimensions, full agreement with findings from Green's function  is established to high order in the expansion, 
 \eq{SSS} -- \eq{SSTTo}. We have  also found  new spectral integral representations \eq{b2nNew}, \eq{eqn:specint} for some of the  local heat kernel coefficients, and explicit results for scalar heat kernels \eq{Spectral_b2n}, applicable in general dimension.  A virtue of the  
 coefficients from Green's functions 
 \eq{bcScalarBar},  \eq{eqn:resvector}, and  \eq{eqn:restensor}  
 is that they can straightforwardly be extended to non-integer dimension, including for the coefficients $c_{d+2n}$, a feat which is much harder to achieve using  spectral sums. 

On a different tack, Green's functions of scalars and transverse vectors on maximally symmetric spaces  are of interest for  applications in cosmology on de Sitter backgrounds, $e.g.$~\cite{Tsamis:2006gj,Miao:2011fc,Miao:2011fc,Mora:2012zi}. As a new addition, we now have derived 
 the Green's function for the Laplacian acting on transverse tensor fields on a fully symmetric background, \eq{GreenTensor} -- \eq{eqn:yensorstructuresolve}. We expect that this will be of use in cosmological settings which are sensitive to the graviton propagator.

Finally, our results are of  practical relevance for a number of farther reaching applications  including  the AdS/CFT correspondence for conformal higher spin models
\cite{Giombi:2013fka,Giombi:2014yra}, effective actions for $N=1$ supergravity \cite{David:2009xg}, or  trace anomalies \cite{Bastianelli:2017wsy}. In quantum gravity,
 possible applications of our results include the perturbative renormalisation of Hořava gravity \cite{Barvinsky:2017mal},
and tests of the asymptotic safety conjecture where intriguing hints for the near-Gaussianity of  gravitational scaling exponents  have  been observed in  \cite{Falls:2013bv,Falls:2014tra,Falls:2016wsa,Falls:2018ylp}.
In the latter, studies thus far have adopted  optimised renormalisation group flows  \cite{Litim:2001up,Litim:2003vp}, which  only depend on  a few leading  heat kernel coefficients   \cite{Codello:2008vh}. Our findings  enable  new  investigations which are sensitive to many more coefficients
  without resorting to flat backgrounds and spectral sums or approximations thereof. 
\\[3ex]

\centerline{\bf Acknowledgments}
\noindent
We thank Basem El-Menoufi and Christoph Rahmede for  discussions. This work  is supported by the Science and Technology Facilities Council (STFC).

\appendix

\renewcommand{\thesection}{{\bf \Alph{section}}}

\section{\bf Expansion Coefficients}
\label{sec:vecstrucfunc}

In the main body, we encounter expansions of vector and tensor structure functions of the form
\begin{equation}\label{expand}
	S (\sigma) = \sum_{n = 0}^\infty S_n \sigma^n \,.
\end{equation}
The expansion coefficients $S_n$ for the transverse vector  Green's function \eq{VectorExpansionCoefficients} where $S\equiv S^T$ and transverse traceless tensor Green's function  \eq{TensorExpansionCoefficients}  where $S\equiv S^{TT}$ are required for the computation of heat kernels, see
\eq{ExpansionVector2} and \eq{TensorContract}, respectively.

The expansion coefficients of the structure function  for transverse vectors are given by
\begin{equation}\label{ST1}
	\begin{split}
		S^T_1 =& \frac{\left(\frac{R}{d (d-1)}\right)^{d/2 - 1}}{(4 \pi)^{d/2} \Gamma \left(\frac{d}{2}\right) (R-d Q)^2}
   \Bigg[ \frac{((d-1) d Q-R)}{(d-1) \sin \left(\frac{\pi  d}{2}\right)} \cos (\pi  \overline{\xi} ) \Gamma \left(\frac{d - 1}{2} + \overline{\xi}\right)\Gamma \left(\frac{d - 1}{2} - \overline{\xi}\right) \\
		& \qquad -\frac{\Gamma (d-1) (d Q-R) \left(\pi  \cot \left(\frac{\pi 
   d}{2}\right)+\psi ^{(0)}(d-1)+\gamma -1\right)}{d-3}+\frac{R \Gamma
   (d-1)}{d-1} \Bigg] \, , \\
		S^T_2 =& \frac{2^{-d-1} \pi ^{-d/2} \left(\frac{R}{d (d-1)}\right)^{d/2 - 1}}{3 (d-1)^2 d^3 (d+2) \Gamma \left(\frac{d}{2}\right) \left(Q-\frac{R}{d}\right)^2} \Bigg[(5 d-8) R^2 \Gamma (d-1) \\
		&+ \frac{((d-1) d Q-R) (3 (d-1) d Q+2 d R-5 R)}{\sin \left(\frac{\pi  d}{2}\right)} \cos (\pi  \overline{\xi} ) \Gamma \left(\frac{d - 1}{2} - \overline{\xi} \right) \Gamma \left(\frac{d - 1}{2} + \overline{\xi}\right) \\
		&-\frac{ R \Gamma (d) (d Q-R) \left(5 d\,\gamma-8 d+\pi  (5 d-8) \cot \left(\frac{\pi  d}{2}\right)+(5 d-8) \psi ^{(0)}(d-1)-8 \gamma +17\right)}{d-3} \Bigg] \, ,
   	\end{split}
\end{equation}
where $\gamma$ denotes the Euler-Mascheroni constant and $\psi^{(0)}(x)\equiv \Gamma'(x)/\Gamma(x)$  the digamma function. We also recall that the parameter $\overline{\xi}$ is given by
\begin{equation}
	\overline{\xi} = \frac{1}{2} \sqrt{d \left(-\frac{4 (d-1) Q}{R}+d-2\right)+5} \,.
\end{equation}
The above expressions are required to obtain the  result \eq{vectorkernel} stated the main text. Note that the dependence of the expansion coefficients \eq{ST1} on the Euler-Mascheroni constant and the digamma function drops out in the final expression \eq{vectorkernel}.

For the expansion of the Green's function for  transverse traceless tensors, we need to calculate the trace of $G_{Q, \mu \nu \rho' \sigma'}^{TT}$  in the coincidence limit, see \eq{GTT}. Using the solution \eq{SolutionTensor} for the Green's function of transverse traceless tensors and the corresponding expansion \eq{expand} for the structure function,
we develop the term $\mathcal{R}_{\mu \rho'} \mathcal{R}_{\nu \sigma'} S^{TT} (\sigma)$  up to order eight in $\sigma_\mu$. We find
\begin{equation}\label{ExpansionTensor8}
	\begin{split}
		&\mathcal{R}_{\mu \rho'} \mathcal{R}_{\nu \sigma'} S^{TT} (\sigma) \\
		=& S^{TT}_0 \Bigg[ \left( \frac{R}{2 (d-1) d} - \frac{R^2 \sigma }{12 (d-1)^2 d^2} + \frac{R^3 \sigma ^2}{180 (d-1)^3 d^3} -\frac{R^4 \sigma ^3}{5040 (d-1)^4 d^4} \right) \sigma _{\mu } \sigma _{\rho '} g_{\nu  \sigma'} \\
		& \qquad + \left( \frac{R}{2 (d-1) d} - \frac{R^2 \sigma }{12 (d-1)^2 d^2} +\frac{R^3 \sigma ^2}{180 (d-1)^3 d^3} - \frac{R^4 \sigma ^3}{5040 (d-1)^4 d^4} \right) g_{\mu  \rho'} \sigma _{\nu } \sigma _{\sigma '} \\
		& \qquad + \left( \frac{R^2}{4 (d-1)^2 d^2} - \frac{R^3 \sigma }{12 (d-1)^3 d^3} + \frac{R^4 \sigma ^2}{80 (d-1)^4 d^4} \right) \sigma _{\mu } \sigma _{\rho '} \sigma _{\nu } \sigma _{\sigma '}+g_{\mu  \rho '} g_{\nu \sigma '}\Bigg] \\
		& + S^{TT}_1 \Bigg[\left( \frac{R \sigma }{2 (d-1) d} - \frac{R^2 \sigma ^2}{12 (d-1)^2 d^2} + \frac{R^3 \sigma ^3}{180 (d-1)^3 d^3} \right) \sigma _{\mu } \sigma _{\rho'} g_{\nu  \sigma '} \\
		& \qquad + \left( \frac{R \sigma }{2 (d-1) d} - \frac{R^2 \sigma ^2}{12 (d-1)^2 d^2} + \frac{R^3 \sigma ^3}{180 (d-1)^3 d^3} \right) g_{\mu  \rho '} \sigma _{\nu } \sigma _{\sigma '} \\
		& \qquad + \left(\frac{R^2 \sigma }{4 (d-1)^2 d^2}-\frac{R^3 \sigma ^2}{12 (d-1)^3 d^3}\right) \sigma _{\mu } \sigma _{\rho '} \sigma _{\nu } \sigma _{\sigma '}+\sigma  g_{\mu  \rho '} g_{\nu  \sigma '}\Bigg] \\
		& +  S^{TT}_2 \Bigg[ \left(\frac{R \sigma ^2}{2 (d-1) d}-\frac{R^2 \sigma ^3}{12 (d-1)^2 d^2}\right) \sigma _{\mu } \sigma _{\rho '} g_{\nu  \sigma '} + \left(\frac{R \sigma ^2}{2 (d-1) d}-\frac{R^2 \sigma ^3}{12 (d-1)^2 d^2}\right) g_{\mu  \rho '} \sigma _{\nu } \sigma _{\sigma '} \\
   & \qquad +\frac{R^2 \sigma ^2}{4 (d-1)^2 d^2} \sigma _{\mu } \sigma _{\rho '} \sigma _{\nu } \sigma _{\sigma '} + \sigma ^2 g_{\mu 
 \rho '} g_{\nu  \sigma '}\Bigg] \\
		& + S^{TT}_3 \Bigg[ \frac{R \sigma ^3}{2 (d-1) d} \sigma _{\mu } \sigma _{\rho '} g_{\nu  \sigma '} +\frac{R \sigma ^3}{2 (d-1) d} g_{\mu  \rho '} \sigma _{\nu } \sigma _{\sigma '} +\sigma ^3 g_{\mu  \rho '} g_{\nu  \sigma '}\Bigg] \\
		& + S^{TT}_4 \sigma ^4 g_{\mu  \rho '} g_{\nu  \sigma '} + \mathcal{O} \left( \sigma_\alpha \right)^9 \, ,
   	\end{split}
\end{equation}
Expressions for the expansion coefficients $S^{TT}_n$ are very long and not given here.
Also, acting with the projectors on this expanded term gives long expressions, which are not shown. 
Similarly to the coefficients \eq{ST1}, we observe that the coefficients $S^{TT}_n$ depend individually on the Euler-Mascheroni constant and the digamma function, whereas the final result  \eq{TensorFinal}  is independent thereof. The above expressions are used to arrive at  the results \eq{TensorContract} and \eq{TensorFinal} in the main text.

\section{\bf Heat Kernels in Even Dimensions}\label{sec:heatkerneleven}
 In this appendix we supply the first five heat kernel coefficients for scalars, vectors and tensors in even dimensions, where we may combine  $b_{2n}^{(s)}$ and $c_{2n}^{(s)}$ into a single coefficient by writing
\beq
	\hat{b}_{2n}^{(i)} =  b_{2n}^{(i)} + c_{2n}^{(i)} 
	\label{eqn:btilde}
\eeq
(recall that $c_{2n}^{(i)}=0$ for any $n<\s0d2$).  Also, since  the scalar heat kernel coefficients $c_{2n}^{(0)}$ vanish identically, we  have $\hat{b}_{2n}^{(0)} = b_{2n}^{(0)}$, giving
\beq\label{Even_0}
	\begin{split}
	\hat{b}_0^{(0)} &= 1\\
\hat{b}_3^{(0)}  &= \frac{1}{6} R \\
\hat{b}_4^{(0)} &= \frac{\left(5 d^2-7 d+6\right) R^2}{360 (d-1) d} \\
\hat{b}_6^{(0)} &= \frac{\left(35 d^4-112 d^3+187 d^2-110 d+96\right) R^3}{45360 (d-1)^2d^2} \\
\hat{b}_8^{(0)} &= \frac{\left(175 d^6-945 d^5+2389 d^4-3111 d^3+3304 d^2-516 d+2160\right)
  R^4}{5443200 (d-1)^3 d^3} \,.
	\end{split}
\eeq
For the transverse vector heat kernels we find
\beq\label{Even_1}
	\begin{split}
		\hat{b}_0^{(1)} =& \, d - 1 \, ,\\
		\hat{b}_2^{(1)} =& \, \frac{\delta _{2,d}}{2} R + \frac{d^2-d-6}{6 d} R \, , \\
		\hat{b}_4^{(1)} =& \left( \frac{\delta _{2,d}}{4} + \frac{\delta _{4,d}}{24} \right) R^2 + \frac{5 d^4-12 d^3-47 d^2-186 d+180}{360 (d-1) d^2} R^2 \, , \\
		\hat{b}_6^{(1)} =& \left( \frac{\delta _{2,d}}{16} + \frac{\delta _{4,d}}{96} + \frac{\delta _{6,d}}{450} \right) R^3 \\
		& + \frac{35 d^6-147 d^5-331 d^4-3825 d^3-676 d^2+10992 d-7560}{45360 (d-1)^2 d^3} R^3 \, , \\
		\hat{b}_8^{(1)} =& \left( \frac{\delta _{2,d}}{96}+\frac{\delta _{4,d}}{768}+\frac{\delta _{6,d}}{2700}+\frac{15 \delta _{8,d}}{175616} \right) R^4 + \frac{R^4}{5443200 (d-1)^3 d^4} \Big(175 d^8-1120 d^7-866 d^6 \\
		& \quad -38260 d^5-31985 d^4+34700 d^3+405996 d^2-627840 d+226800 \Big) \, .
	\end{split}
\eeq
The transverse traceless tensor heat kernels give
\beq\label{Even_2}
	\begin{split}
		\hat{b}_0^{(2)} =& \, \frac{(d-2) (d+1)}{2} \, ,\\
		\hat{b}_2^{(2)} =& \, 3 \delta _{2,d} R + \frac{d^3-2 d^2-13 d-10}{12 (d-1)} R \, , \\
		\hat{b}_4^{(2)} =& \left( 3 \delta _{2,d}+\frac{5}{8} \delta _{4,d} \right) R^2 + \frac{5 d^5-17 d^4-105 d^3-475 d^2-620 d-228}{720 (d-1)^2 d} R^2 \, , \\
		\hat{b}_6^{(2)} =& \left(\frac{3 \delta _{2,d}}{2}+\frac{5 \delta _{4,d}}{36}+\frac{14 \delta _{6,d}}{225}\right) R^3 \\
		& +\frac{35 d^7-182 d^6-884 d^5-8618 d^4-21515 d^3-23648 d^2-38116
   d-28032}{90720 (d-1)^3 d^2} R^3 \, , \\
		\hat{b}_8^{(2)} =& \left( \frac{\delta _{2,d}}{2} + \frac{5 \delta _{4,d}}{288} + \frac{7 \delta _{6,d}}{1125} + \frac{675 \delta _{8,d}}{175616} R^4 \right) + \frac{R^4}{10886400 (d-1)^4 d^3} \Big(175 d^9-1295 d^8-4296 d^7 \\
		& \quad -80514 d^6-263073 d^5-709635 d^4-907534 d^3-940876 d^2-2454072 d-1896480 \Big) \, .
	\end{split}
\eeq
Results can now be compared  with  \cite{Rahmede:2008dwa,Falls:2017lst} where expressions have been given  for the heat kernel coefficients $\hat{b}_{2n}$ in even dimensions. 
In $d = 4$, our findings for $\hat{b}_{2i}^{(n)}$ agree numerically with the corresponding expressions  $b_{2i} |_n$ given in \cite{Rahmede:2008dwa} and in appendix $B$ of \cite{Falls:2017lst}, except for $b_8 |_1$. For general $d$,  deviations appear in
the algebraic expressions for $b_8 |_1$ and   $b_8 |_2$,  and 
some contributions which uniquely arise in even integer dimensions (proportional to Kronecker deltas) have been  missed. Our results are  consistent with the heat kernel, spectral sums, and expressions for general backgrounds as found in the literature.

\section{\bf Heat Kernels from Spectral Integrals}
\label{app:scalarspec}

In this appendix we detail the derivation of spectral integrals and heat kernel coefficients from spectral sums in general dimension, using dimensional continuation as outlined in \sct{sct:anacon}. We focus on scalars for simplicity. The computational steps for transverse vectors and transverse traceless tensors are the same and only differ by the specific form of the multiplicities.

We start by taking  the dimension $d$ to be non-integer. Results for integer dimension then follow from dimensional continuation. Basic input are the eigenvalues and their multiplicities  (see Tab.~\ref{tab:eigvmulti}), which for scalars are given by
\beq\label{C1}
	D_\ell^0 = \frac{(2 \ell + d - 1) \Gamma \left( \ell + d - 1 \right)}{\Gamma (d) \Gamma (\ell + 1)}\,.
\eeq
A large $\ell$ expansion is performed using the asymptotic series \cite{GammaQasym}
\beq
	\frac{\Gamma \left( \ell + a \right)}{\Gamma \left( \ell + b \right)} \simeq \ell^{a - b} \sum_{n = 0}^\infty \frac{1}{\ell^n} 
	{a - b \choose n} \,B_n^{(a - b + 1)} (a) \,,
\eeq
where the ``$\simeq$''  indicates the  asymptotic nature of the series. The generalised Bernoulli polynomials $B_n^{(\ell)} (x)$ are defined  as the Taylor coefficients of
\beq\label{Bnlx}
 \left( \frac{t}{e^t - 1} \right)^\ell e^{x t} = \sum_{n = 0}^\infty B_n^{(\ell)} (x) \frac{t^n}{n!} \,.
\eeq
For the particular case \eq{C1}, this gives us
\beq
	D_\ell^0 \simeq \frac{2 \ell + d - 1}{\Gamma (d)} \sum_{n = 0}^\infty \ell^{d - 2 - n} {d - 2 \choose n} B_n^{(d - 1)} (d - 1) \, .
	\label{Dasymexp}
\eeq
We now have to perform the spectral integral
\beq
	\int_0^\infty \text{d} \ell \, D_\ell^0 e^{- \lambda_\ell^0 t} \simeq \frac{1}{\Gamma(d)} \sum_{n = 0}^\infty {d - 2 \choose n} B_n^{(d - 1)} (d - 1) \int_0^\infty \text{d} \ell \, \ell^{d - 2 - n} (2 \ell + d - 1) e^{- \frac{\ell (\ell + d - 1)}{d (d - 1)} R t} \, ,
\eeq
which can be done term by term with the help of
\beq
	\begin{split}
		\int_0^\infty \text{d} \ell \, \ell^{d-1-n} e^{- \frac{\ell (\ell + d - 1)}{d (d - 1)} R\, t} =& 
		\0{1}{2} \bigg(\0{R\, t}{d (d - 1)}\bigg)^{-d/2} \,
		\bigg[ I_{d,n}(R\,t)-(d-1) J_{d,n}(R\,t)\bigg]\,,
	\end{split}
	\label{lint}
\eeq
where we have introduced the shorthand notations
\beq
\label{IJ}
	\begin{split}
		I_{d,n}(x)&=
		\bigg(\0{x}{d (d - 1)}\bigg)^{n/2} \,\Gamma \bigg( \0{d-n}{2}\bigg) \, _1F_1 \bigg( \0{d-n}{2}, \0{1}{2}, \0{d - 1}{4 d}  x\bigg) 
		\\ 
		J_{d,n}(x)&=
		\left(\0{x}{d (d - 1)}\right)^{(n + 1)/2} \, \Gamma \bigg(\0{d-n+1}{2} \bigg) \,_1F_1 \bigg(\0{d-n+1}{2}, \0{3}{2}, \0{d - 1}{4 d} x \bigg)  
\end{split}
\eeq
 involving the Kummer function  ${}_1F_1(a,b,z)=\sum_{k=0}^\infty\frac{a_{(k)}}{b_{(k)}}\frac{z^k}{k!}$, and $a_{(k)}$ the rising factorial. The expression \eq{lint} with \eq{IJ} is  valid for general $d > 1$, $R\, t > 0$, and $n < d$. By analytic continuation in the dimension we extend its domain of validity to all $n$. The right-hand side of \eq{lint}  contains poles   from  Gamma functions provided $d - n$ is a negative integer or zero. However, these poles will be  multiplied by zeros from the binomial coefficient
 in \eq{Dasymexp}, giving a finite result for any $d - n$ if analytically continued from non-integer $d$.
Following this strategy, we find
\beq
	\begin{split}
		\left.\int_0^\infty \text{d} \ell \, D_\ell^0 e^{- \lambda_\ell^0 t} \right|_{\rm reg.}=& 
		\0{2}{\Gamma(d)} \left( \frac{d (d - 1)}{R\,t} \right)^{d/2}
\sum_{n = 0}^\infty {d - 2 \choose n} B_n^{(d - 1)} (d - 1) \times \\
		& \bigg[ \s0{1}{2} I_{d,n}
		(R \, t) 
		- \s0{d-1}{2}  J_{d,n}(R \, t) 
		+ \s0{d - 1}{4}  I_{d,n+1}(R \, t) 
		- \s0{(d-1)^2}{4} J_{d,n+1}(R \, t) 
		\bigg] \,,
	\end{split}
	\label{specintscres}
\eeq
where "reg." indicates our procedure of analytic continuation and the usage of an asymptotic large $\ell$ expansion.  Note that  the right-hand side  contains terms proportional to $t^{-d/2+m}$ and  $t^{-(d+2m+1)/2}$ for any integer $m\ge 0$. The former relate to the heat kernel coefficients $b_{d+2m}$. The latter cannot arise from a heat kernel expansion, and, hence, their coefficients must vanish identically.  With this in mind, we  write \eq{specintscres} as 
\beq
	\begin{split}
		\left.\int_0^\infty \text{d} \ell \, D_\ell^0 e^{- \lambda_\ell^0 t} \right|_{\rm reg.}=& 
		\0{1}{\Gamma(d)} \left( \0{d (d - 1)}{R\,t} \right)^{d/2}\times \\ & 
		\left\{\ \ \,
		\sum_{n = 0}^\infty 
		{d - 2 \choose 2n}
		B_{2n}^{(d - 1)} (d - 1) 
		\left[I_{d,2n}(R\,t)-\0{1}{2}(d-1)^2\,J_{d,2n+1}(R\,t)\right]\right.
		\\ & 
		\left. 
		+\0{d-1}2 \sum_{n = 0}^\infty
		 {d - 2 \choose 2n + 1}
		B_{2n + 1}^{(d - 1)} (d - 1) \bigg[I_{d,2n}(R\,t)-2J_{d,2n+1}(R\,t)\bigg]\right\}
	\end{split}
	\label{specintscresbpart}
\eeq
which manifestly contains only terms of the form $\sim t^{-d/2+m}$. Together with \eq{eqn:specint} this gives rise to the result \eq{Spectral_b2n} in the main text.

As a final  check, we confirm that any terms of the form $t^{-(d+2m+1)/2}$ in \eq{specintscres} vanish, as they must. Their sum can be written as
\beq\label{consistency}
	\begin{split}
		0 =&   \0{d-1}{2} \sum_{n = 0}^\infty {d - 2 \choose 2n} B_{2n}^{(d - 1)} (d - 1)
		 \bigg[  I_{d,2n+1}(R\,t)-2 J_{d,2n}(R\,t)\bigg]
\\
		& + \sum_{n = 0}^\infty {d - 2 \choose 2n + 1} B_{2n + 1}^{(d - 1)} (d - 1)
		  \bigg[ I_{d,2n+1}(R\,t) -\012 (d-1)^2\, J_{d,2n}(R\,t)\bigg]\,,
	\end{split}
\eeq
and we have checked the validity of \eq{consistency}  for general dimension through a series expansion in curvature $R$ up to $\mathcal{O} \left( R^{10} \right)$. We expect this to hold true to any order.

The same steps which have led to \eq{specintscresbpart} can be repeated for transverse vectors and transverse traceless tensors leading to corresponding expressions.
\newpage

\bibliography{bibtex}
\bibliographystyle{mystyle}

\end{document}